\documentclass[twocolumn]{aastex63}

\usepackage{xfrac}
\usepackage{float}
\usepackage{physics}
\usepackage{amssymb}
\usepackage{savesym}
\savesymbol{tablenum}
\usepackage{rotating}
\usepackage[section]{placeins}
\usepackage[noabbrev]{cleveref}
\usepackage{subfigure}
\usepackage{gensymb}
\usepackage{siunitx}
\usepackage{mathtools}

\restoresymbol{SIX}{tablenum}
\newcommand{\tus}[1]{$_{\text{2}}$}

\newcommand{\dg}{\degree}

\shorttitle{MEGSASIM: Earth Trojan Lifetimes}
\shortauthors{Yeager et al.}

\graphicspath{{./}{figures/}}

\begin{document}

\title{MEGASIM: Lifetimes and Resonances of Earth Trojan Asteroids - The Death of Primordial ETAs?}

\correspondingauthor{Travis Yeager}
\email{yeagerastro@gmail.com}

\author[0000-0002-2582-0190]{Travis Yeager}
\affiliation{Lawrence Livermore National Laboratory \\
7000 East Ave, Livermore \\
Livermore, CA 94550, USA}

\author[0000-0002-2582-0190]{Nathan Golovich}
\affiliation{Lawrence Livermore National Laboratory \\
7000 East Ave, Livermore \\
Livermore, CA 94550, USA}

\begin{abstract}
We present an analysis of lifetimes and resonances of Earth Trojan Asteroids (ETAs) in the MEGASIM data set \citep{Yeager_2022}. Trojan asteroids co-orbit the Sun with a planet but remain bound to the Lagrange points, L4 (60° leading the planet) or L5 (60° trailing). In the circular three-body approximation, the stability of a Trojan asteroid depends on the ratio of the host planet mass and the central mass. For the inner planets, the range of stability becomes increasingly small, so perturbations from the planets have made primordial Trojans rare. To date there have been just two ETAs (2010 TK\textsubscript{7} and 2020 XL\textsubscript{5}), several Mars Trojans, and a Venus Trojan discovered. The estimated lifetimes of the known inner system Trojans are less than a million years, suggesting they are interlopers rather than members of a stable and long-lasting population. With the largest ETA n-body simulation to date, we are able to track their survival across a wide initialized parameter space. We find the remaining fraction of ETAs over time is well fit with a stretched exponential function that when extrapolated beyond our simulation run time predicts zero ETAs by 2.33 Gyr. We also show correlations between ETA ejections and the periods of the Milankovitch cycles. Though Earth's orbital dynamics dominate the instabilities of ETAs, we provide evidence that ETA ejections are linked to resonances found in the variation of the orbital elements of many, if not all of the planets.
\end{abstract}



\section{Introduction}

\indent Asteroids that co-orbit with a planet near its fourth or fifth Lagrange points (hereafter L4 and L5) are called Trojan asteroids\footnote{`Trojan' has generally been used to describe objects co-orbiting with Jupiter; however, given the general nature of these minor planets across the Solar System, we use the term `Trojan' generically and refer to which planet they co-orbit throughout.}. L4 and L5 are stable in the reduced three-body problem for all the planets in the Solar System. Nearly ten thousand are associated with Jupiter's L4 and L5 points as of 2021, and there are expected to be roughly equal 1 km sized Jupiter Trojans to the number in the main asteroid belt \citep{2005AJ....130.2900Y}. Trojans have been observed co-orbiting with Venus, Earth, Mars, Jupiter, Uranus and Neptune \citep{2000DPS....32.1407C, 2009A&A...508.1021A, 2010MPEC....T...45G, 2011Natur.475..481C, 2013MNRAS.432L..31D, 2015MNRAS.453.1288D, 2017RNAAS...1....3D, 2018LPI....49.1771Y, 2021RNAAS...5...29D}.

\indent To date, there are two known Earth Trojan asteroids (hereafter ETAs): $2010 TK_7$ and $2020 XL_5$ \citep{2000DPS....32.1407C,  2011Natur.475..481C, 2022NatAs...6..178H}, which both librate far from the Lagrange points within the space between Earth and L3 on trajectories known as `tadpole orbits'. The two ETAs that have been observed are not in highly stable regimes and are expected to remain ETAs on a time scale of only tens of thousands of years, as determined by numerical simulations \citep{2012A&A...541A.127D, 2021RNAAS...5...29D, 2022NatCo..13..447S}. If libration amplitudes remain small ETAs can remain bound to L4 or L5, though if perturbed enough the ETAs can cross near the L3 Lagrange point, and continue to co-orbit but on the opposite side of the Solar system, which is then classified as a horseshoe orbit. Further perturbations will eventually lead to ejection from the co-orbiting regime entirely. In the restricted three body problem, \citet{1999ssd..book.....M} derived analytical constraints for the maximum radial distance an ETA can travel from Earth's orbit and still remain in the tadpole potential well:

\begin{equation}\label{eq:1}
    \partial r \le \partial r_{crit} = \sqrt{\frac{8\mu_2}{3}}
\end{equation}
where $r_{crit}$ is the radial distance of the ETA from the host orbit, and $\mu_2=M_\Earth / M_\Sun$. For the Earth, $r_{crit}$ is approximately 0.0011 au. The two known  ETAs, $2010 TK_7$ and $2020 XL_5$, have heliocentric semi-major axis determined to be a = $0.999273055\pm8\times10^{-9}$ AU and $1.0009 \pm 0.0002$ AU, respectively \citep{2021RNAAS...5...29D}. Both of these semi-major axis currently are within $1 \pm r_{crit}$ criteria for the Earth, but even these are stable for a short period of time compared with the age of the Solar System.

\indent There are a few possible sourcing mechanisms for ETAs. They could have formed in the proto-planetary disk and fallen into the eventual L4 or L5 \citep{2020A&A...642A.224M}. At a slightly later stage, they may have been captured debris produced from major planetary impacts or planetary migration. Finally, they may be sourced from perturbed populations of asteroids elsewhere in the Solar System such as ejected asteroids from the main asteroid belt. The first two options would have occurred early in the formation of the Solar System, and we consider these to be ``primordial'' in nature. Primordial sourcing through the impact hypothesis may also explain lunar formation. The giant impact hypothesis, where the Earth was impacted by another early planet that came from a 1 AU orbit in the solar nebula has been proposed. \citet{2005AJ....129.1724B} suggest a Mars-sized proto-planet could have formed in a stable orbit among debris at either Earth's L4 or L5 Lagrange point and collided with Earth to form the eventual Earth--Moon system. It is reasonable to expect a cloud of debris from such a collision may have provided a primordial source of ETAs \citep{2012Sci...338.1052C}. If found, such a population of asteroids would offer a unique lens into the evolution of the early Solar System and the Earth and Moon system. This fact has driven much interest in their discovery. 

\indent Modeling of the ETA population with n-body simulations first appeared in the literature by \citet{1967AJ.....72....7S}, with results showing there to be regions where long term stability of ETAs can exist. Since then a plethora of numerical simulations have be carried out, but our focus is on a recent simulations that study the lifetimes of ETAs. \citet{2010DPS....42.1303J} modeled the lifetimes and stability of three distinct populations of ETAs, built by allowing only one of the semi-major axis, eccentricity, or inclination to vary from that of Earth's orbit. These populations were defined as the following: `A', whose semi-major axes were varied between 0.988 and 1.012 AU, `E' with eccentricities ranging between 0 and 0.4, and the `I' population with inclinations varied between 0 and 180\degree{}. The `A' population of ETAs was found to retain 80\% of the initial population by 100 Myr.  The `E' and `I' ETA populations were found to be much more unstable as a whole, with the `E' population only retaining 20\% of the ETAs by 100 Myr, and the `I' population falling to less than 10\%.  The `I' population, unlike the other two populations, was also found to lose approximately half of the ETAs suddenly after 1 Myr of simulation time. 

\indent A potential population of ETAs is suggested by \citet{Cuk2012MNRAS.426.3051C}, who analyzed the long term stability of both tadpole and horseshoe orbits in a sample of 1000 asteroid orbits. The initial semi-major axis are allowed to be one of 25 values evenly spaced from 1 to 1.01 AU with 40 possible inclinations evenly spaced from 0 to 40 degrees. They found a fraction of their initial set of asteroids remained stable as Earth horseshoes and tadpole orbits at the end of their 700 Myr Earth simulation. The orbits that remained stable until 1 Gyr were found to exist with semi-major axis between 1 and 1.002 AU, eccentricities between 0 and 0.2 and inclinations from 0\degree{} to 19\degree.

\indent \citet{2013CeMDA.117...91M} applied a frequency map analysis \citep[e.g.,][]{marzari2002saturn} to an ETA simulation which indicated some ETA orbits are Gyr-stable . In this case, the prediction of long term stability is gleaned from a \SI{e5} year integration. A frequency map analysis in Keplerian orbital element parameter space was applied to a 12 Myr data set \citep{Zhuo2019A&A...622A..97Z}. Two stability regions were found for inclinations below $15\degree{}$ and another $24\degree{}$ to $37\degree{}$. The orbital instabilities for inclinations between 15\dg{} and 24\dg{} are believed to arise from the $\nu_{3}$ and $\nu_{4}$ secular resonances. Nodal secular resonance $\nu_{13}$ is able to change initial inclinations by up to 10\dg{} and the $\nu_{14}$ resonance could increase inclination 20\dg{} or more. \citet{Zhuo2019A&A...622A..97Z} found that tadpole orbits experience great differences in the frequency spaces of their secular precession and that secular precession in tadpole orbits are more sensitive to the frequency drift of the inner planets. This could indicate a clearance of all tadpole orbits. Furthermore, an asymmetry was found between the prograde and retrograde rotating ETAs possibly due to the Yarkovsky effect \citep[for a historical reference see][]{2005JBAA..115..207B} -- a thermal effect in which the light from the Sun heats up the asteroids, the re-radiated light causes a small perturbation radially, which dependent upon spin, will result in an inward or outward migration relative to the Sun. This effect produces an increase in orbital instability for small asteroids.  If an ETA is to survive 1 Gyr they find it must have a radius larger than 90 m for a prograde spin and 130 m for a retrograde spin. For an ETA to survive the age of the Solar System a radius of 278 m for a prograde spin and 400 m for a retrograde spin is required.

\indent Recently, \citet{2021MNRAS.507.1640C} found similar results to \citet{Cuk2012MNRAS.426.3051C, Zhuo2019A&A...622A..97Z} insofar as the stability of horseshoes is found to be greater than tadpoles, and that horseshoes and tadpoles may easily survive the age of the Solar System. From the simulation data a constraint of 3-12 ETAs with no more than 60 may exist bound to the L4 and L5 points. The simulated tad poles were found to have excitation in the orbital eccentricities that persist until the eccentricity is beyond 0.1 or the semi-major axis deviates from that of Earth by more than 0.0028 au. Horseshoes were instead found to escape through an excitation of the orbital inclination for angles greater than 13\degree{}. The simulations provide further evidence to that in \citet{2012A&A...541A.127D, Zhuo2019A&A...622A..97Z}, that secular resonances are responsible for destabilising ETAs. The fraction of ETAs that remain over the simulation appears to fall roughly linearly from 0 to 750 Myr, flattens until 1.3 Gyr before continuing down at a slightly shallower linear rate until the end of the 2 Gyr simulation.

\indent Large N-body simulations that map the full picture of ETA stability and trajectories are important in placing limits on observational attempts to detect ETAs. \citet{2000Icar..145...33W} found the densest location of the ETA distribution to be slightly inside the orbit of Earth near L4 and determined observability of their simulated ETA populations. \citet{2012MNRAS.420L..28T} further explored optimal search strategies. A case was made that 70-80 ETAs are missing from observations \citep{2019DDA....5010006C}. 

\indent Observers have searched for such primordial populations suggested by the simulations. \citet{2021AJ....161..282L} used DECam data and a novel shift-and-stack method to set a tight upper limit on ETAs with H$<$21.77. They find their observations support an ETA population at L4 of $N\textsubscript{ETA} < 1$ for $H = 14.28$, $N\textsubscript{ETA} < 7$ for $H = 16$, and $N\textsubscript{ETA} < 642$ for $H = 22$. They also reviewed and updated the upper limits of all other known searches for this population, at L4 \citep{2018LPI....49.1771Y} and the similar L5 \citep{1998Icar..136..154W, 2018LPI....49.1149C, 2020MNRAS.492.6105M}. Though these results eliminate the chances of a large population of 1 km-class ETAs at L4, they do not eliminate a population of $\sim$100 m ETAs, which remain a target of deep interest for Solar System formation and planetary defense. 

Here we present results of the highest fidelity simulations of ETAs to date \citep{Yeager_2022}. The MEGASIM offers a complete coverage of the initial Keplerian orbital element parameter space, where as previous work generally fixed multiple parameters to be equal to Earth, the MEGASIM requires no such conditions on the 6-dimensions required to initialize ETA orbits. The analysis in this paper centers on the lifetimes (survivability) with a specific interest in answering whether there can exist a primordial population of such objects. In \S\ref{sec:model} we describe our model. In \S\ref{sec:results} we present our analysis and results, and in \S\ref{sec:conclusions} we offer a discussion of our results compared with the literature and our conclusions.

\section{The Model}\label{sec:model}

\indent We utilize two different integrators to build two distinct data sets. One set of simulations was carried out using the IAS15 n-body integrator and another using the Wisdom-Holman symplectic integrator referred to from here on as WHfast \citep{wh,reboundwhfast}. Both integrators are a part of the \texttt{REBOUND} package \citep{rebound,reboundias15}. The Solar Systems are initialized with the Sun and eight planets, Mercury, Venus, Earth-Moon, Mars, Jupiter, Saturn, Uranus and Neptune. The initial planetary positions are gathered from the JPL HORIZONS system at an epoch of JD2458475.30035 \citep{1996DPS....28.2504G}. We utilized the Catalyst, Quartz and Ruby clusters managed by Livermore Computing at Lawrence Livermore National Laboratory \footnote{\url{https://hpc.llnl.gov/hardware/platforms}}.

\indent The two ETA data sets in this paper will be referred to by the integrator that was used (IAS15 and WHfast, respectively). Initially only the IAS15 integrator was used to generate the orbits of ETAs and was chosen because of its ability to use a variable time step meaning near approaches to Earth would be well-resolved. While the IAS15 integrator lends to very high accuracy it comes with a high computational cost, limiting the duration simulations could reach. The choice was then made to carry out another set of simulations with the WHfast integrator to reach much longer time scales. The \texttt{REBOUND} WHfast integrator takes several parameters to define its use, a time step of 1 day, symplectic correctors were set to 17th order (\texttt{sim.ri\_whfast.corrector} = 17) and the \texttt{sim.ri\_whfast.safe\_mode} parameter was turned off. Runs were initially conducted without the use of sympletic correctors, and it was found that the stable orbits of the ETAs was poorly resolved. This comes with a benefit of being able to compare the results from both integrators using large number statistics to better understand any possible limitations of using the WHFast integrator.

\indent Parallelization of the simulation is achieved by running 44,800 initializing batches of 500 asteroids injected with six orbital parameters selected from a random distribution, as described below. The number 500 was chosen to balance running enough asteroids so that individual simulations would have asteroids in them without wasting CPU time integrating mostly planets and running few enough asteroids so that the 24 hr wall clock limit for the high-performance computing (HPC) clusters was not prohibitive of simulation progress. Each model is then integrated forward with \texttt{REBOUND}. Position and velocity information for each is output every 1000 years but the integration time steps are shorter. The IAS15 simulations were carried out until either no stable ETAs remained or until a simulation of 50 million years was achieved. The set of simulations integrated with WHFast are not cut off until they either have lost all stable  ETAs or until a simulation time of 1 Gyr was reached.

\indent Removal of ETAs occurs if they drift beyond the Earth -- L3 line, effectively removing any horseshoe or non-co-orbiting objects. Asteroids that were ejected due to complete instability eventually crossed this line and were removed.

\begin{figure}
    \centering
    \includegraphics[width=\columnwidth]{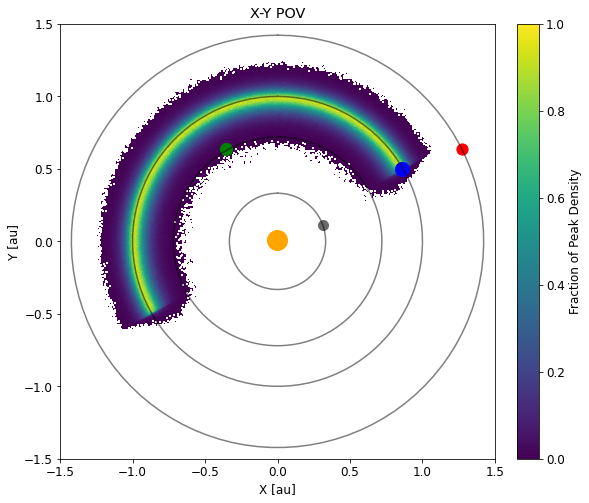}
    \caption{The initialization of the WHfast simulation with all 11.2 million initial ETAs show as a heatmap. The Sun, Mercury, Venus, Earth and Mars are represented as the orange, grey, green, blue and red dots. The circles are provided for reference and do not represent the exact orbits of the planets.}
    \label{fig:initialization}
\end{figure}

\indent The ETA orbits are initialized with a true longitude ($\theta$) randomly and uniformly distributed between Earth and L3 on the side of Earth's orbit containing L4. The argument of pericenter and longitude of ascending node are randomly assigned values between 0 and $2\pi$ assuming a uniform distribution. The semi-major axis ($a$), eccentricity ($e>0$) and inclination ($i$) were sampled from a three dimensional Gaussian with mean and co-variance

\begin{equation}
    \boldsymbol{\mu} = \begin{pmatrix} 
    \bar{a} \\
    \bar{e} \\
    \bar{i}
    \end{pmatrix} = \begin{pmatrix} 
    1\,\text{au} \\
    0 \\
    0\degree{}
    \end{pmatrix} ~
    \boldsymbol{\Sigma} = \begin{pmatrix} 
    0.025\,\text{au} & 0 & 0 \\
    0 & 0.075 & 0 \\
    0 & 0 & 15\degree{}   
    \end{pmatrix}.
\end{equation} A visualization of the 11.2 million initialized orbits is provided in \Cref{fig:initialization}. The initial positions of the inner planets is also provided, with circular pseudo-orbits overlaid for reference.

\indent The initial Keplerian orbital elements extended well beyond ranges expected for ETAs. Importantly, none of the orbital elements are fixed to that of Earth, which physically may not be required for stability. This is a markedly different setup from the simulations in the literature, which all fixed a subset of the orbital elements and often varied only one a time. While our initial parameters included many orbits we are not interested in, our method for removing asteroids from the simulation does not allow any Earth co-orbits that cross the Earth-Sun-L3 line to remain in the simulation for very long. For a horseshoe orbit to exist within the simulation, it would have to find itself on the L4 side of the Earth-Sun-L3 line every 1000 years to avoid removal from the simulation. This is exceedingly unlikely, and we took care to remove those that did survive a few timesteps by chance from our results. Our broad initialization does result in a computationally inefficient simulation, but in the end it results in the broadest look at ETA stability at high statistical fidelity over a Gyr, which is truly unprecedented in the literature. 

\section{Results}\label{sec:results}

\subsection{Comparing the two simulations}
\label{sec:comparing_sims}

\begin{figure}
    \centering
    \includegraphics[width=\columnwidth]{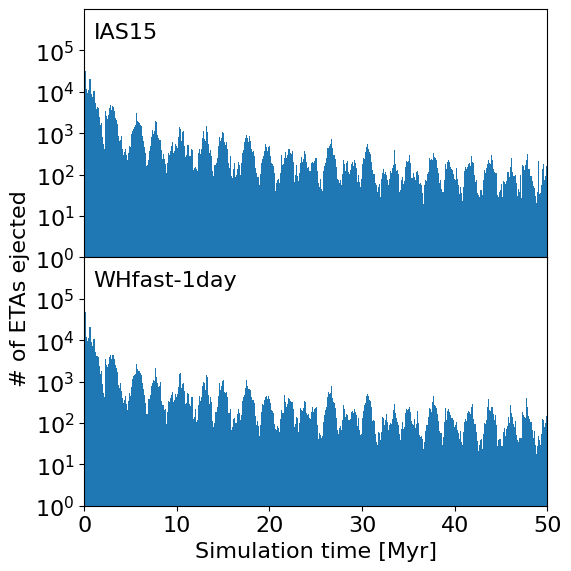}
    \caption{Binned ETA ejections over time for the two simulations. Bin widths are 100 kyr. The top panel shows the IAS15 simulation which was only integrated for 50 Myr. The bottom panel shows the first 50 Myr of the WHfast simulation for comparison.}
    \label{fig:lifetimes_binned}
\end{figure}

\indent  \Cref{fig:lifetimes_binned} shows a histogram of binned ejections over time for the first 50 Myr of each simulation. The loss of asteroids is very similar for both integrators. The height of the bins corresponds to the number of ETAs that are removed from the simulation with bin widths of one hundred thousand years. There is a periodic oscillation in the number of ETAs removed with time, apparent from the height of the bins with a period of roughly 2 Myr. This finding is discussed further in \S\ref{subsec:ejections}.

\begin{figure}
    \centering
    \subfigure[]{
    \includegraphics[width=\columnwidth]{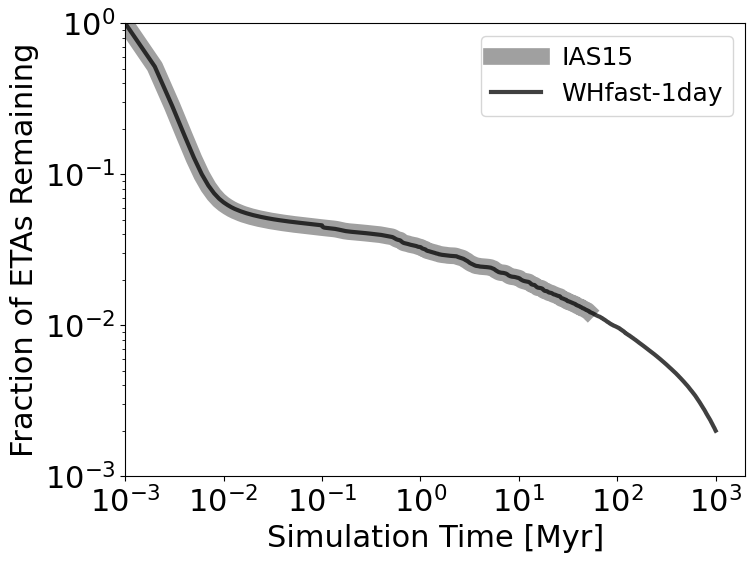}
    }
    \subfigure[]{
    \includegraphics[width=\columnwidth]{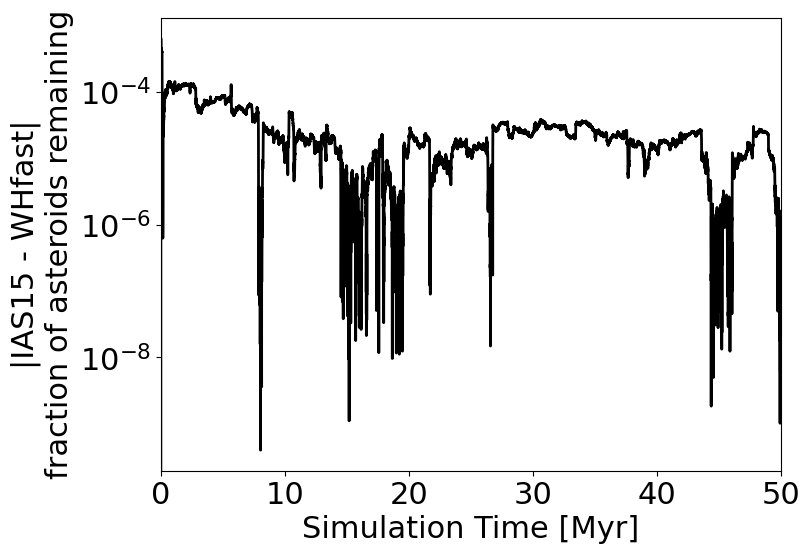}
    }
    \caption{Panel (a) IAS15 (thick gray) and WHfast (thin black) simulation fraction of ETAs remaining over time.  The IAS15 stops at 50 Myr and the WHfast simulation continues to 1 Gyr. Panel (b) is the absolute value of the difference between the two curves in panel (a). For reference, the initial population of ETAs in both sims is 11.2 million.}
    \label{fig:simdata}
\end{figure}

\indent The simulations begin with 11.2 million ETAs and after every simulation time step (1000 yr) the number of ETAs remaining in the simulation is recorded. The resulting curve, shown in panel (a) of \Cref{fig:simdata}, is the number of remaining ETAs counted every 1000 years divided by the number initialized. An ETA lifetime is then defined as the amount of time the ETA spent in the simulation before being removed due to crossing the plane perpendicular to the ecliptic that contains the Earth and L3; that is, asteroids co-orbiting with Earth but leading by no more than 180$\degree$.

In the early stages of the simulation the fraction of remaining ETAs rapidly decreases to fewer than 10\% within the first $\sim$10 kyr. This initial phase of ejections is indicative of our broad initialization. We earlier drew a distinction between these loosely bound interlopers\footnote{There is value in studying these objects as they closely resemble the two known ETAs. Future work will study this early ejected population.} and the long term ETAs that we seek to study here. After 10 kyr, the ejection rate slows for the majority of the simulation.

The absolute difference between the two integrators (IAS15 and WHfast) remaining ETA fractions is provided in panel (b) of Figure \ref{fig:simdata}. The typical difference between the two integrated data sets ETA fractions is of order \SI{e-5}, which corresponds to $\sim$100 out of 11.2 million ETA particle difference between the sims. We initially worried that the use of the fixed time step integrator for WHFast would lead to significant differences in long term stability estimation as the effect of near Earth approaches may have been smoothed over. These results show that the WHfast integrator works extremely well at matching the results of the much more computationally intensive IAS15 integrator for timescales of 50 Myr. The deviation between the two models remains very consistent even at 50 Myr indicating the accuracy is maintained on significantly longer time scales. The presence of the periodic ejections in both panels of Figure \ref{fig:lifetimes_binned} further suggests that the two simulations are capturing the same physics to high fidelity. Upon finding this, we cut off integration for the IAS15 simulation at 50 Myr and carried forward the WHFast simulation alone. The remainder of our results will focus on the WHFast results.

\subsection{Ejections from the WHfast data set}\label{subsec:ejections}

\subsubsection{Short Term Ejection Signatures}\label{subsubsec:short}
\begin{figure*}
    \centering
    \subfigure[]{
        \includegraphics[width=1\textwidth]{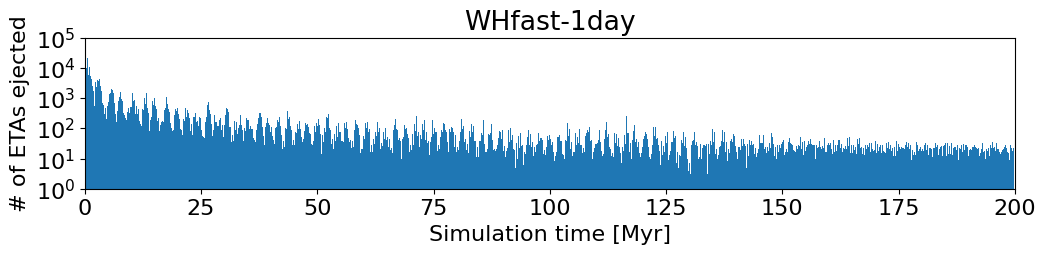}
    }
    \subfigure[]{
    \includegraphics[width=1\textwidth]{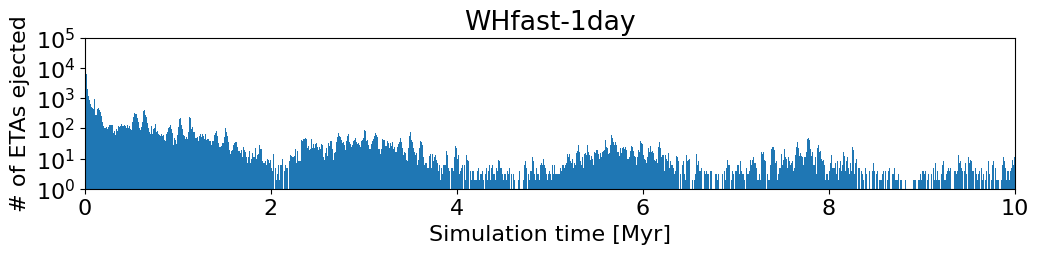}
    }
    \caption{Panel (a) shows the binned ETA ejections over time spanning 0 to 200 Myr of the WHfast data set. Bin widths are 100 kyr the same as \Cref{fig:lifetimes_binned}. Panel (b) shows the binned ETA ejections over time spanning 0 to 10 Myr of the WHfast data set. Bin widths are 1000 yr.}
    \label{fig:whfastejections}
\end{figure*}

\indent The periodic ejection pattern apparent in \ref{fig:lifetimes_binned} has a large amplitude. Ejections vary by as much as a factor of ten with a rough period of $\sim$2 Myr. Panel (a) of \Cref{fig:whfastejections} shows this continues for WHFast to beyond 100 Myr. At 150 Myr the periodicity begins to wash out and by 200 Myr it no longer apparent. Panel (b) of \Cref{fig:whfastejections} shows this periodicity at higher time-resolution with 1000 year bins. There are higher frequency oscillations apparent as well indicating a blend of ejection sources. 

\begin{figure}
    \centering
    \hspace*{-.5cm}
    \includegraphics[width=\columnwidth]{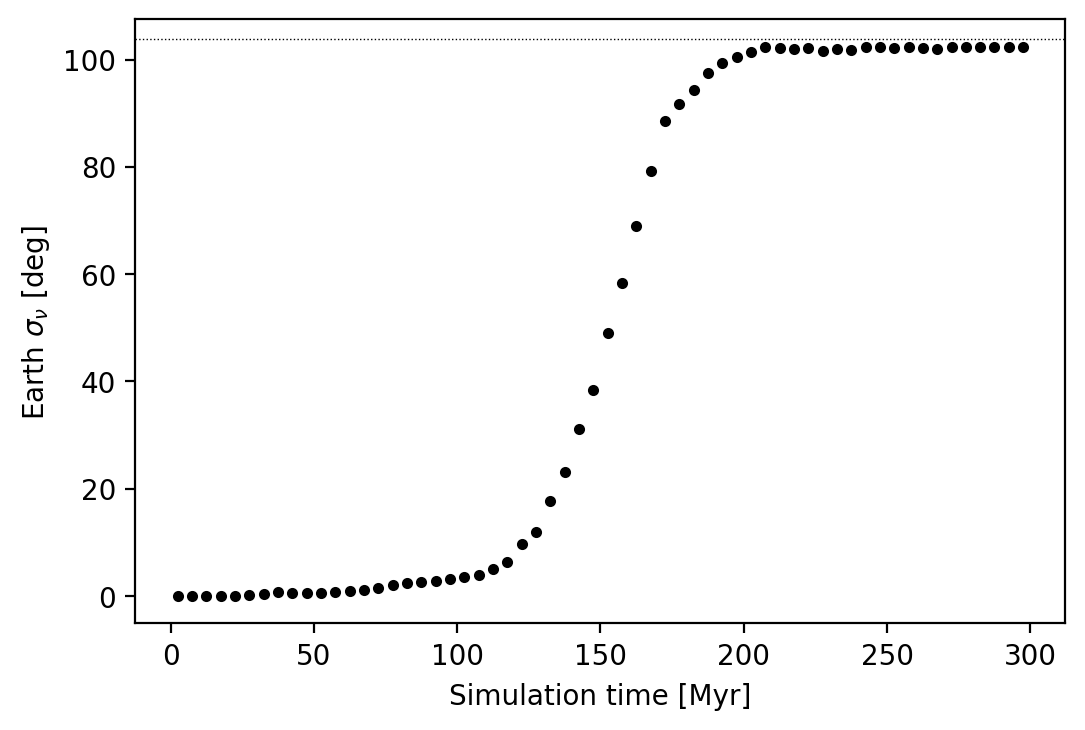}
    \caption{The standard deviation of the true anomalies in Earth in all 22,400 simulations as a function of simulation time. In the first 100 Myr, the simulations are largely in agreement, but between 100 and 200 Myr, the numerical drift of the simulations spreads out the individual Solar Systems. By 200 Myr, the standard deviation approaches the value for the standard deviation of a uniform distribution spanning 360$^{\circ}$ ($\sqrt{(360-0)^2/12}\approx103.92$). This value is shown with a horizontal dotted line. This behavior of the member simulations matches very well with the decoherence of the oscillatory ejection behavior present in panel (a) of Figure \ref{fig:whfastejections}.}
    \label{fig:numericaldrift}
\end{figure}

Here we explore the evolution of the individual parallel simulations relative to one another. In order to simulate so many asteroids, we initialized 22,400 identical Solar Systems each with 500 ETAs. \Cref{fig:numericaldrift} shows how the position of the planets in each simulation deviate from one another over time. The 100-200 Myr time-frame matches well with the diminishing of the periodic ejection pattern apparent in Figure \ref{fig:lifetimes_binned}. This demonstrates that the oscillatory behavior is likely always present; however, it is the coherence of the individual Solar Systems diminishing rather than the periodicity to cause the cumulative binned ejections to change appearance beyond 150 Myr. 

\begin{figure*}
    \centering
    \subfigure[]{
    \includegraphics[width=1\textwidth]{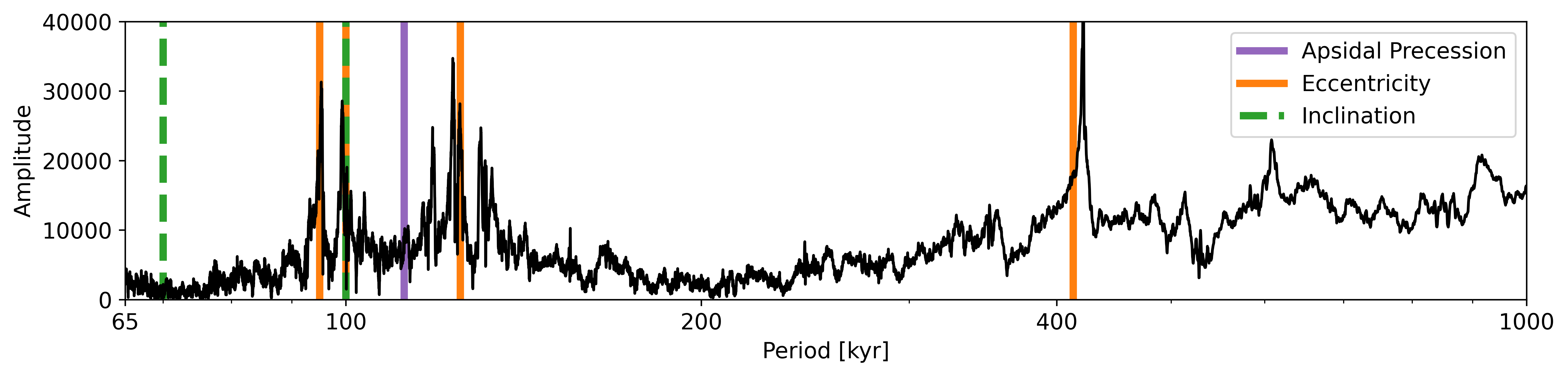}
    }
    \subfigure[]{
    \includegraphics[width=1\textwidth]{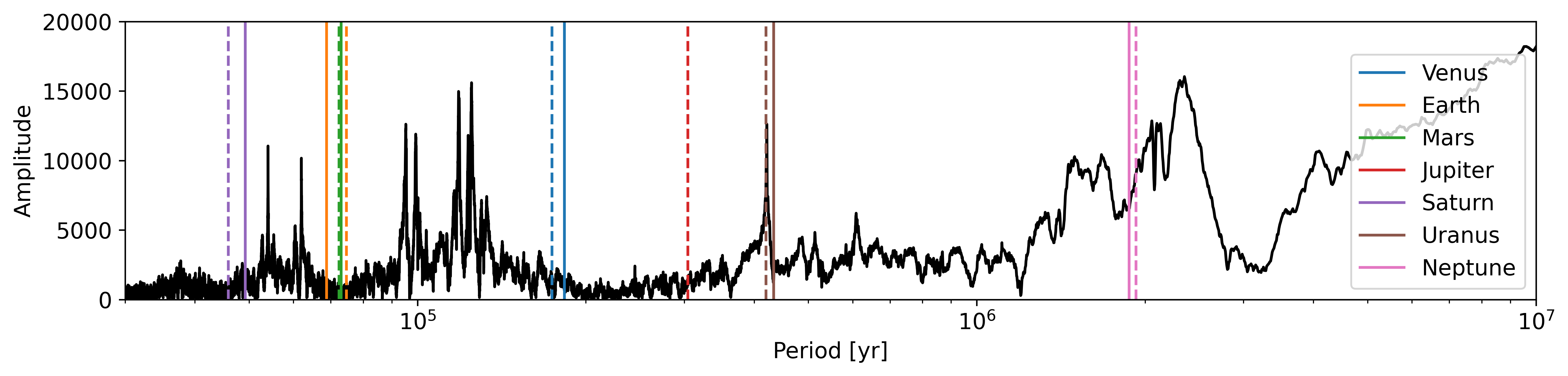}
    }
    \caption{The black line is the ETA ejection FFT spectrum. The x-axis is the period of a signal in years and the y-axis is the relative strength of a given signal. Panel (a) spans 65 kyr to 1 Myr with the color lines representing the periods of the Milankovitch cycles. The orange lines are the periods of known modes in the Milankovitch cycles due to changes in Earth's eccentricity at 95, 125 and 413 kyr, which loosely combine to a period of 100 kyr. In green are periods of inclination changes at 70 and 100 kyr. Finally in purple is the period of apsidal precession at 112 kyr. Panel (b) shows the positions of the planetary $s$ (solid) and $g$ (dashed) resonances overlaid on the ETA ejection FFT, spanning 10 kyr to 10 Myr. Color indicates the corresponding planet, listed in the legend on the right.}
    \label{fig:resonance_milan}
\end{figure*}

\indent In \Cref{fig:resonance_milan} we present a Fast Fourier Transform (FFT) of the ETA ejections with a number of important frequencies. The FFT is computed from the bin counts of a histogram of ETA lifetimes with bin widths of 1000 years. In the top panel of Figure \ref{fig:resonance_milan}, we show the alignment of several peaks in the 65 kyr to 1 Myr range with the well-known Milankovitch cycles, which were first considered to describe changes in Earth's climate and associated with variations in the orbit of Earth. The periods for known changes to Earth's eccentricity, inclination and pericenter are overlaid on the ETA ejection FFT. Periods due to changes in eccentricity are shown as orange vertical lines at 95, 125 and 413 kyr, which loosely combine to a period of 100 kyr, green lines are periods of inclination changes at 70 and 100 kyr, and the purple line is the period of apsidal precession at 112 kyr \citep{muller1997spectrum,2011A&A...532A..89L}. It is not surprising that Earth's orbital dynamics are a driving factor in ETA ejections, but to our knowledge this is the first indication that the same variations in Earth's orbit can be linked to both climate dynamics and ETA dynamics. 

\indent In the bottom panel of Figure \ref{fig:resonance_milan}, we compare the ejection FFT with $s$ and $g$ resonances of each planet. The periods of the planetary $s$ (solid) and $g$ (dashed) resonances overlaid on the ETA ejection FFT, spanning 10 kyr to 10 Myr are colored by planet. The resonances for the inner planets are from \citet{laskar1990chaotic} and the outer planets \citet{1989A&A...210..313N}. The $s$ and $g$ resonances of the planets apart from Uranus do not directly match up with any peaks in the ETA ejection FFT. The resonances of Neptune may play a role in the ETA ejections as they are in the range of the multiple ETA ejection peaks centered around 2.3 Myr that we discussed above. The $s$ resonance of Jupiter is found at 129 Myr off the right of panel (b), but there is no correlated peak in the ETA ejection FFT, so we truncated the figure.

\begin{figure*}
    \centering
    \includegraphics[width=1\textwidth]{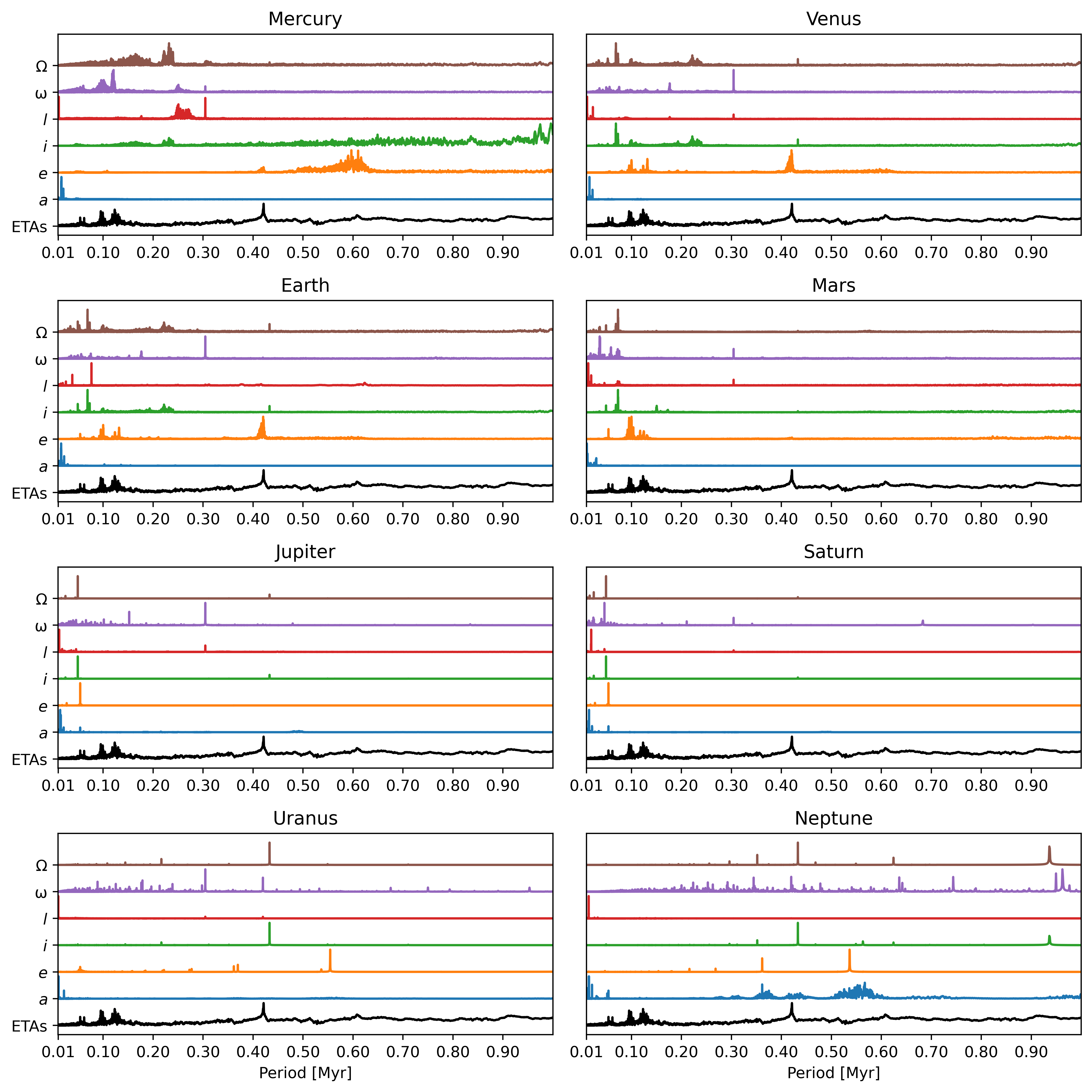}
    \caption{FFT of planetary orbital elements using 1 Myr to 1 Gyr of simulation data. Panels only span 10 kyr to 1 Myr to focus on the many lower frequencies that exist in both the planetary orbital elements and ETA ejections. In each panel is a different planet with line colors from bottom up: black/ETA ejections, blue/semi-major axis, orange/eccentricity, green/inclination, red/true longitude, purple/argument of periapsis, brown/longitude of the ascending node.}
    \label{fig:fft_kyr}
\end{figure*}

\indent Next we compare the FFTs of all orbital elements for each planet and the ETA ejections. The planet orbital element FFTs are computed using a single Solar System simulation spanning 1 Myr to 1 Gyr. We checked for consistency across a handful of our parallel Solar Systems and found good qualitative agreement. \Cref{fig:fft_kyr} shows the FFT for six Keplerian orbital elements for each planet (separated by subplot) along with the FFT of the ETA ejections for comparison. The vertical axis labels indicate the Keplerian orbital element, but for further guidance, they are from top to bottom: longitude of the ascending node (brown), argument of periapsis (purple), true longitude (red), inclination (green), eccentricity (orange) and semi-major axis (blue). At the bottom of each panel, we show the ETA ejection FFT for comparison (black). The vertical axis is scaled such the the highest power for a given FFT is of equal height -- i.e., the relative heights between the various FFTs should not be compared. 

\indent The Milankovitch cycles shown in \Cref{fig:resonance_milan} appear in the orbital FFTs of Earth's orbital elements.  Earth's eccentricity shows the three modes near periods of 95, 125 and 413 kyr and the inclination at 70 kyr, though the 100 kyr period is a very small peak. The simulation orbital element data FFTs correspond well to the spread of frequencies seen in ejection FFT (black). However, the peak with a period just larger than 400 kyr lines up not only with Earth's eccentricity mode, but also strong signal in Venus's eccentricity FFT as well as Uranus's and Neptune's inclination FFT.

\begin{figure*}
    \centering
    \includegraphics[width=1\textwidth]{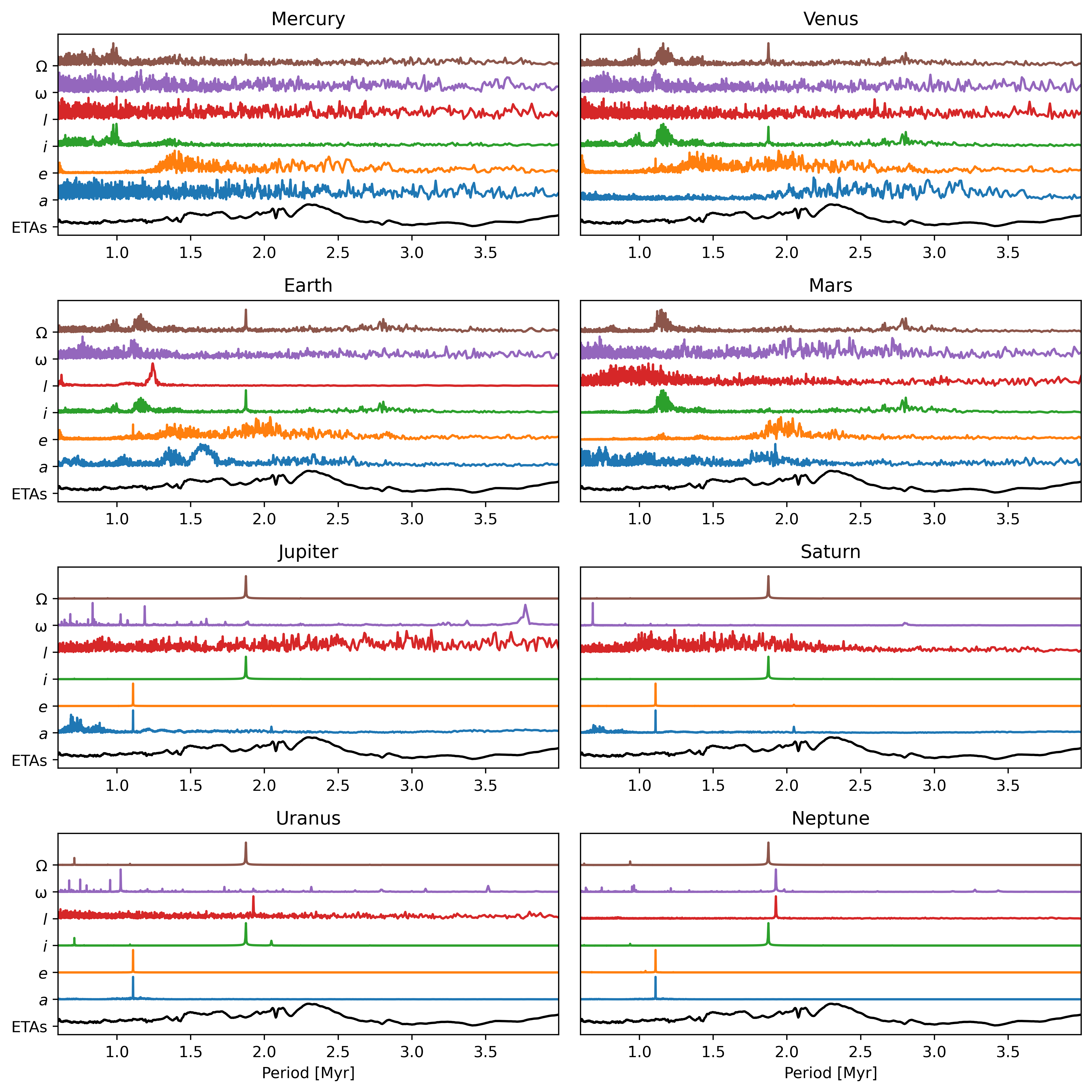}
    \caption{Same as \Cref{fig:fft_kyr} except highlighting periods in the range of 1 to 4 Myr.}
    \label{fig:fft_myr}
\end{figure*}

\Cref{fig:fft_myr} is presented analogously to \Cref{fig:fft_kyr} for FFT signals with longer periods. The longest period of ejection oscillation peaks near 2.3 Myr with the gradual increase in FFT power from 1.5 Myr to 2.75 Myr. This $\sim 2$ Myr period corresponds to the oscillation seen in Figure \ref{fig:lifetimes_binned} and \ref{fig:whfastejections}. The full structure of the ETA ejection FFT seems to be built from an additive combination of many planetary resonances, involving all components of planetary orbital variation.

\subsubsection{Long Term Ejection Signatures}\label{subsubsec:long}

\indent Previous work found fall off rates of initialized ETAs of various functional forms including logarithmic and and linear functions \citep{2010DPS....42.1303J, Cuk2012MNRAS.426.3051C, 2013CeMDA.117...91M, Zhuo2019A&A...622A..97Z, 2021MNRAS.507.1640C}. Our simulation is different in that we initialized a much broader region of orbital space, and we simulated many more asteroids out to 1 Gyr. In the literature, the ETAs were initialized on elliptical orbits with precise Earth-like properties. For example, in previous simulations the longitude of the ascending node and argument of perihelion have not been varied from that of Earth. Initializing ETA orbits over the full Keplerian parameter space provides a full accounting of the zones of stability over long time-scales, the orbital elements from which they are possibly initialized, and their associated lifetimes. It is our primary goal to study this broader population of potential ETAs.

\begin{figure}[htb]
    \centering
    \includegraphics[width=\columnwidth]{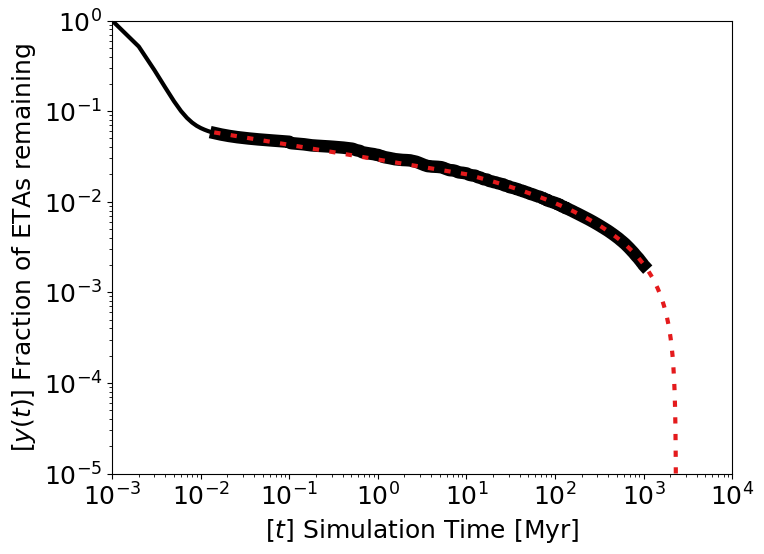}
    \caption{The remaining ETA fraction as a function of simulation time in black with a best fit stretched exponential (red dashed line) overlaid. y(t) and t refer to variables found in \Cref{eq:stretchexp}. The area where the black line is thicker is used for the stretched exponential fit to avoid the initial stages of the simulation where very unstable ETAs filter out. The fit spans 15 kyr to 1 Gyr.}
\label{fig:functionfits}
\end{figure}

\indent We consider here the fractional remaining population as a function of time at the time resolution of our simulation output (1000 year increments). For the first $\sim$10 time steps, we find a logarithmic fall off as the most unstable initialized orbits are quickly cleared out (see the leftmost portion of the black curve in the top panel of \Cref{fig:simdata}). By 15 kyr the rate of change in the surviving fraction of ETAs departs from the rapid logarithmic fall off and is well modeled by a stretched exponential from 15 kyr to 1 Gyr. This is a function of the form\begin{equation}
    y(t) = A \exp(\lambda t^\beta) + C,
    \label{eq:stretchexp}
\end{equation}where A, $\lambda$, $\beta$ and C are free parameters. We show the fit with a red dashed curve in \Cref{fig:functionfits}. The black line is the WHFast simulation data with the thick black portion indicating the time span for which the function was fit. The fit, when extrapolated beyond the simulation time predicts zero asteroids (we will call this time $t_0$) by $t_0=2.33$ Gyr. We will discuss the implications of this in \S\ref{sec:conclusions}. 

\indent To assess the uncertainty in our stretched exponential fit and extrapolated time to zero remaining asteroids, we preformed a bootstrap re-sampling analysis on the asteroid lifetimes. For each initialized asteroid, we stored the time it was either ejected or 1 Gyr for those that survive the duration. To create bootstrap samples of the data, we selected 11.2 million asteroid lifetimes from our data (with replacement) and fit a stretched exponential to the realized fractional remainder as a function of time. For the full data set, our bootstrap analysis gives a mean and 80\% confidence interval of $t_0 = 2.33^{+3.04\times10^{-3}}_{-4.55\times10^{-4}}$ Gyr. Thus our simulation provides evidence that a primordial population of ETAs that were leftover from the Solar System formation (i.e., that were around when the planets assumed their current arrangement as our simulation was initialized) would not survive to the current age of the Solar System.

\begin{figure}[htb]
    \centering
    \includegraphics[width=\columnwidth]{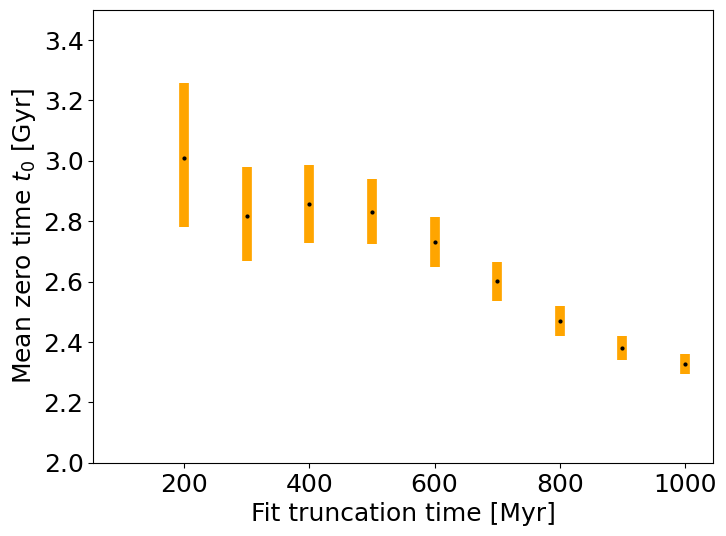}
    \caption{The range of  predicted by the stretched exponential fits to bootstrap resampled data as the domain over which the fit is performed is increased. The fits all begin at 15 kyr and extend to 100 to 1 Gyr incremented by 100 Myr. The width of the orange lines extends from the 10th to 90th percentile obtained from bootstrap samples of remaining ETA distribution.}
\label{fig:bootstrap}
\end{figure}

\indent Since our extrapolated lifetime is less than the age of the Solar System, we performed an analysis on the estimated zero time ($t_0$) value of a stretched exponential fit as a function of the amount of our simulation data used. That is, we fit a stretched exponential to our bootstrap samples that are truncated at 100 Myr intervals out to 1 Gyr. For all fits, we have fit from 15 kyr to the truncated time. The results of this procedure are presented in \Cref{fig:bootstrap}. The black dot indicates the mean $t_0$ (vertical axis) predicted for all curves fit to a given range of data, and the orange bars span the $80\%$ confidence interval for the bootstrap samples. As more data is fit, the stretched exponential tightens to a $t_0$ under the age of the Solar System. We will discuss this further in \S\ref{sec:conclusions}. The mean and 80\% confidence intervals for all parameters of Equation \ref{eq:stretchexp} and $t_0$ for the bootstrap analysis are presented in \Cref{table:fitparameters}. 

\indent Finally, $\beta$ in \Cref{eq:stretchexp} can be physically interpreted as one over the averaged relaxation time of the relaxing species, in our case the entire ETA population. Based on the best fit $\beta$ values of \Cref{table:fitparameters}, a corresponding relaxation time of the ETAs increases with time, ranging from 6.6 Gyr with a 100 Myr data fit up to 8.8 Gyr when fitting to the full 1 Gyr of data. Such a long relaxation time in the ETA population would indicate that chaos will eventually lead to a depletion of all ETAs, as our stretched-exponential fit suggests. The perturbations to the ETA orbits cause both short and long timescale diffusion of the ETAs. Thus they can never truly reach an equilibrium before further perturbations occur.

\section{Discussion}\label{sec:conclusions}

\begin{table*}[htb]
\movetableright=-0.95in
\resizebox{\textwidth}{!}{%
\begin{tabular}{|c|c|c|c|c|c|}
\hline
$t_{\text{end}}$ [Myr] & $A$ & $\lambda$ & $\beta$ & $C$ & $t_0$ [Gyr] \\
\hline
100 & $0.096^{+1.95\times10^{-3}}_{-1.46\times10^{-3}}$ & $-1.0284^{+1.47\times10^{-2}}_{-1.04\times10^{-2}}$ & $0.161^{+3.35\times10^{-3}}_{-2.64\times10^{-3}}$ & $-0.001^{+2.46\times10^{-4}}_{-1.87\times10^{-4}}$ & $7.20^{+7.29\times10^{-1}}_{-3.37\times10^{-1}}$ \\
200 & $0.110^{+2.02\times10^{-3}}_{-4.20\times10^{-3}}$ & $-1.1098^{+1.36\times10^{-2}}_{-2.31\times10^{-2}}$ & $0.141^{+2.49\times10^{-3}}_{-5.42\times10^{-3}}$ & $-0.004^{+2.13\times10^{-4}}_{-5.90\times10^{-4}}$ & $3.01^{+7.62\times10^{-2}}_{-3.61\times10^{-1}}$ \\
300 & $0.113^{+1.46\times10^{-3}}_{-3.35\times10^{-3}}$ & $-1.1239^{+1.01\times10^{-2}}_{-1.74\times10^{-2}}$ & $0.138^{+1.73\times10^{-3}}_{-4.15\times10^{-3}}$ & $-0.004^{+1.38\times10^{-4}}_{-4.79\times10^{-4}}$ & $2.82^{+2.70\times10^{-2}}_{-2.30\times10^{-1}}$ \\
400 & $0.112^{+3.05\times10^{-4}}_{-1.74\times10^{-3}}$ & $-1.1189^{+3.25\times10^{-3}}_{-7.58\times10^{-3}}$ & $0.139^{+4.65\times10^{-4}}_{-2.37\times10^{-3}}$ & $-0.004^{+8.88\times10^{-6}}_{-3.05\times10^{-4}}$ & $2.86^{+3.01\times10^{-2}}_{-1.46\times10^{-1}}$ \\
500 & $0.113^{+1.12\times10^{-4}}_{-6.76\times10^{-5}}$ & $-1.1227^{+6.15\times10^{-4}}_{-3.16\times10^{-3}}$ & $0.138^{+1.93\times10^{-5}}_{-5.44\times10^{-4}}$ & $-0.004^{+3.24\times10^{-5}}_{-1.43\times10^{-4}}$ & $2.83^{+4.50\times10^{-2}}_{-6.78\times10^{-2}}$ \\
600 & $0.116^{+4.04\times10^{-4}}_{-1.21\times10^{-3}}$ & $-1.1429^{+3.94\times10^{-3}}_{-1.13\times10^{-2}}$ & $0.135^{+5.24\times10^{-4}}_{-7.04\times10^{-4}}$ & $-0.004^{+1.11\times10^{-5}}_{-3.72\times10^{-5}}$ & $2.73^{+2.66\times10^{-2}}_{-2.27\times10^{-2}}$ \\
700 & $0.121^{+6.68\times10^{-4}}_{-2.06\times10^{-3}}$ & $-1.1789^{+5.47\times10^{-3}}_{-1.64\times10^{-2}}$ & $0.130^{+7.02\times10^{-4}}_{-1.33\times10^{-3}}$ & $-0.005^{+2.81\times10^{-5}}_{-1.78\times10^{-5}}$ & $2.60^{+2.02\times10^{-2}}_{-5.54\times10^{-3}}$ \\
800 & $0.130^{+1.71\times10^{-3}}_{-3.01\times10^{-3}}$ & $-1.2305^{+1.15\times10^{-2}}_{-2.14\times10^{-2}}$ & $0.123^{+1.40\times10^{-3}}_{-1.80\times10^{-3}}$ & $-0.005^{+9.43\times10^{-5}}_{-6.37\times10^{-5}}$ & $2.47^{+6.14\times10^{-3}}_{-3.20\times10^{-3}}$ \\
900 & $0.138^{+2.72\times10^{-3}}_{-3.23\times10^{-3}}$ & $-1.2792^{+1.67\times10^{-2}}_{-2.19\times10^{-2}}$ & $0.117^{+1.90\times10^{-3}}_{-1.69\times10^{-3}}$ & $-0.006^{+1.48\times10^{-4}}_{-5.89\times10^{-5}}$ & $2.38^{+8.85\times10^{-4}}_{-9.06\times10^{-4}}$ \\
1000& $0.145^{+3.41\times10^{-3}}_{-3.53\times10^{-3}}$ & $-1.3164^{+1.98\times10^{-2}}_{-2.29\times10^{-2}}$ & $0.113^{+2.13\times10^{-3}}_{-1.68\times10^{-3}}$ & $-0.006^{+1.77\times10^{-4}}_{-6.16\times10^{-5}}$ & $2.33^{+3.04\times10^{-3}}_{-4.55\times10^{-4}}$ \\
\hline
\end{tabular}}
\caption{The mean and 80 $\%$ confidence intervals for the stretched exponential parameters and the extrapolated time to zero remaining ETAs. $t_{\text{end}}$ gives the truncated time over which the fits were performed. All fits begin at 15 kyr. $A$, $\lambda$, $\beta$, and $C$ are the parameters of the stretched exponential of form: \Cref{eq:stretchexp}.  $t_0$ is when each curve crosses 0 on the y-axis. Values provided are the mean of the bootstrap population with their 80$\%$ confidence intervals. The confidence intervals span the 10--90 quantile range for the bootstrap samples of our ETA simulated lifetime data. \label{table:fitparameters}}
\end{table*}

\subsection{The physics of the stretched exponential}
\label{sec:stretched_exponential}

\indent The number of ETAs remaining with time is found to follow a stretched exponential form. Here we explore the literature of stretched exponential functions, which have been used frequently in scenarios similar to ours. \citet{DOBROVOLSKIS2007481} fit the survival of simulated particles ejected from the Saturnian Moon Hyperion with a stretched exponential, and also compiled a list of many other related astrophysics particle removal simulations where the stretched exponential appears \citep{gladman1990fates, holman1993dynamical, levison1994long, gladman1995dynamical, dones1996dynamical, gladman1996exchange, farinella1997disruption, gladman1997destination, holman1997possible, burns1998dynamically, dones1999dynamical, evans1999possible, malyshkin1999keplerian, dobrovolskis2000dynamical, gladman2000near, hartmann2000time, tabachnik2000asteroids, robutel2001frequency, armstrong2002rummaging, chambers2002symplectic, nesvorny2002long, quintana2002terrestrial, nesvorny2003orbital, dobrovolskis2004fate, alvarellos2005fates, gladman2005impact, zeehandelaar2005material, hamilton2006fate, zahnle2007transfer, alvarellos2008transfer}. 

\indent Our further literature review found stretched exponential used to model physical processes in many different areas of research, with the common thread being underlying chaotic dynamics. We provide here an incomplete but expansive list to demonstrate the breadth of the literature:
\begin{itemize}
  \setlength\itemsep{0em}
    \item General disordered systems \citep{2019PhDT.......109L}
    \item Li$^7$ nuclear spin-lattice relaxation \citep{trinkl20007, kaps2001heavy, johnston2005dynamics}
    \item Chaos in the formation of cosmological large scale structure \citep{2015arXiv151001909B}
    \item Cosmological decay of Higgs-like scalars \citep{2019PhRvD.100b3531B}
    \item Thermal relaxation of self gravitating sheet model \citep{2010JSMTE..10..012J}
    \item Solar flare distributions \citep{2020ApJ...894...66N}
    \item Star-forming cloud models \citep{2012ApJ...750...13C}
    \item Sand drift wind models \citep{2020E&PSL.54416373Y}
    \item Viscous remnant magnetization dating method applied to tsunami boulders \citep{2019E&PSL.520...94S}
    \item Buoyancy driven turbulence \citep{bershadskii2016buoyancy}
    \item Distributed chaos and isotropic turbulence \citep{bershadskii2015distributed} 
    \item Modeling of earthquake aftershocks \citep{2009PEPI..175..183G, 2015GeoRL..42.9726M}
    \item Rainfall distribution \citep{wilson2005fundamental}
    \item Solid-state physics for creep, annealing, electrical impedance, and magnetic spin relaxation \citep{palmer1984models,peterson1989time,halsey1991stretched,scher1991time,phillips1996stretched} 
\end{itemize}

\indent Based on the literature review above, it is not surprising that our ETA remainder fraction is well fit by a stretched exponential, but what is the origin of the mathematical form? 

\indent Here we draw a connection between our findings and a simplified model of the diffusion processes. \citet{DOBROVOLSKIS2007481} suggested diffusion in the space of orbital elements was related to their findings regarding the ejected material from Hyperion. Simplifying from the stretched exponential, we first consider the one-dimensional diffusion equation:
\begin{equation}
    \frac{\partial^2 P(x,t)}{\partial x^2} = \frac{1}{D}\frac{\partial P(x,t)}{\partial t}
    \label{eq:1ddiffusion}
\end{equation}
where $P(x,t)$ is distribution of the diffusing fluid or particles. $D$ is the diffusion constant and incorporates the physics involved with a given diffusion process. This can be separated into spatial and time components ($X$ and $T$, respectively):
\begin{equation}
    \frac{d^2X(x)}{dx^2} = -k^2X(x)
\end{equation}
\begin{equation}
    \frac{dT(t)}{dt} = -k^2DT(t)
\end{equation}
which have solutions:
\begin{equation} 
    X(x) = a \cos{kx} + b \sin{kx}
    \label{eq:diffsol}
\end{equation} 
\begin{equation}
    T(t) = c \exp{-k^2Dt},
\end{equation}respectively, where $a$, $b$, and $c$ are constants related to the boundary conditions, and $k$ is the separation constant. The time-dependence is exponential in nature, with a characteristic relaxation time related to the diffusion constant $D$ and the separation constant $k$. 

\indent Continuing with the idea that the ejection of ETAs is analogous to diffusion processes, let us assume some function $f(t)$ is equal to a series of exponential functions:
\begin{equation}
    f(t) = \sum_{n=0}^{n} C_n \exp{a_n t}
    \label{eq:sumexp}
\end{equation}
where $C_n$ and $a_n$ are for now just some constants for fitting. The philosophy here is that, following \citet{DOBROVOLSKIS2007481}, the diffusion processes in the space of orbital elements was related to the stretched exponential functional form. We know that orbital elements in the many-body gravitational problem can induce resonances that perturb minor planets \citep[e.g., the Kirkwood Gaps in the main asteroid belt][]{1891SidM...10..194K}. However, there are also subtler resonances that occur at non-regular intervals that perturb the ETA population with individual diffusion processes. The derivatives of this formulation is then:
\begin{equation}
    f^{(n)}(t) = \sum_{n=0}^{n} C_n {a_n}^n\exp{a_n t}
    \label{eq:sumexpderivatives}
\end{equation}
The matrix form of this equation is:
\begin{equation}
    \begin{bmatrix}
    f(t)\\
    f'(t)\\
    \vdots\\
    f^{(n)}(t)
    \end{bmatrix} 
    = 
    \begin{bmatrix}
    {a_0}^0 & {a_1}^0 & \dots & {a_n}^0 \\
    {a_0}^1 & {a_1}^1 & \dots & {a_n}^1 \\
    \vdots & \vdots & \ddots & \vdots  \\
    {a_0}^n & {a_1}^n & \dots & {a_n}^n
    \end{bmatrix} 
    \begin{bmatrix}
    C_0 \exp{a_n t}\\
    C_1 \exp{a_n t}\\
    \vdots\\
    C_n  \exp{a_n t}
    \end{bmatrix}
    \label{eq:matrix}
\end{equation}
where the square matrix is known as the Vandermonde matrix, which itself has many interesting connections in applied mathematics. Relevant for our discussion here is the connection between the Vandermonde matrix and the discrete Fourier transform matrix \citep{6951993}. Furthermore, the Vandermonde matrix appears in the theory of Dyson Brownian motion \citep{10.1214/EJP.v13-539}\footnote{We recommend  \url{https://terrytao.wordpress.com/2010/01/18/254a-notes-3b-brownian-motion-and-dyson-brownian-motion/}, an excellent blog post on this topic.}. The relationship to Brownian motion here is especially interesting as it implies the diffusion of ETAs out of the loosely bound potential is associated with a random walk due to accumulating perturbations. These perturbations are caused by consecutive resonances with the planets over long timescales. Brownian motion is often invoked as a random walk in a flat potential, but a potential is not excluded mathematically, and so we interpret our ETA ejection process as each ETA experiencing an accumulation of random perturbations, which are linked with independent diffusion processes with their own characteristic relaxation time. The ETAs in our initial population are broad, and often easily ejected from a random accumulation of these perturbations. Others withstand the random perturbations for long timescales; however, over time the accumulation tends to all the ETAs by a finite timescale.

\indent It has also been shown that a stretched exponential can be represented as a continuous spectrum of pure exponential functions \citep{johnston2006stretched}:
\begin{equation}
    \exp{-\lambda^*t^\beta} = \int_{0}^{\infty} P(s,\beta) \exp{-s t} \,ds
    \label{eq:continuousexp}
\end{equation}where $s\equiv \lambda^*/\lambda$ and $\lambda^*$ is the characteristic relaxation time for a given pure exponential and $P(s,\beta)$ is a probability distribution such that $\int_{0}^{\infty} P(s,\beta) \,ds = 1$. In this formulation, the parameter $\beta$ of \Cref{eq:stretchexp} is related to the logarithmic full-width at half maximum (FWHM) of the probability distribution $P(s,\beta)$. $P(s,\beta)$ has a cutoff at low $s$ that decreases monotonically with decreasing $\beta$ such that $\beta$ can measure the small relaxation-rate cutoff of $P(s,\beta)$. 

\indent Our finding that there are no simple discrete set of resonances from the planets in Figure \ref{fig:resonance_milan}, \ref{fig:fft_kyr}, and \ref{fig:fft_myr} to describe the complicated nature of the ETA ejection FFT further bolsters our discussion here, where we tie the ETA ejections to Brownian motion. The random walk nature of Brownian motion is analogous in some sense here to the accumulation of random perturbations experienced by the ETAs over time. As we showed in \S\ref{subsubsec:long}, the $\beta$ value from our stretched-exponential fit suggests they can never reach equilibrium from prior perturbations before future ones occur.

\subsection{The Yarkovsky Effect}\label{subsec:yarkovsky}

\indent Because we do not account for additional forces beyond n-body gravity the lifetimes of the ETA orbits may be considered as upper bounds for stability. This further strengthens the argument for no existing primordial population of ETAs today. The strength of the Yarkovsky effect and thermal radiation from the Sun depends on the rotation speed, size, albedo and density of a given asteroid. \citet{Zhuo2019A&A...622A..97Z} found smaller asteroids with radii below 90 m or 130 m (depending on if the orbit is rotating prograde or retrograde, respectively) are cleared out of the ETA regime on timescales less then 1 Gyr. This finding suggests that any primordial population of ETAs is likely composed of asteroids larger than 100 meters.

\subsection{The primordial population of ETAs}\label{subsec:primordial}
\indent We would like to point out here that we are particularly focused on primordial ETAs, but it is worth exploring exactly what that means. While it is easy to invoke ETAs coming from the early Solar System, some consideration is needed to think through the implications for the results in this paper.

\indent Since we initialized our asteroids with Earth-like orbits in a Solar System with present day planets, we either have implicitly assumed that asteroids survived migration of the gas giants or that an additional mechanism for sourcing these asteroids occurs after migration. It certainly seems reasonable to assume that very early Solar System ETAs may have been disrupted by gas-giant migration since our findings show that much smaller perturbations eventually can clear out ETAs anyway, so it is a reasonable step to assume much more rapid large scale motion by the planets likely had an out-sized effect on the ETA population.

\indent A single model of the Solar System that can account for each of gas-giant formation, Jupiter Trojan asteroids, the asteroid belt, the Kuiper Belt, the present arrangement of the inner and outer planets, as well as the observed impact histories of the inner planets and the Moon does not yet exist. Several models have succeeded in modeling a subset of the above. One leading model, colloquially named the Nice model (after the Southern French city) provides an explanation for the arrangement of the gas-giants, a means of stripping the proto-planetary disk of its gas in order to generate rocky planets \citep{2005Natur.435..459T}, the distribution of asteroids from the asteroid belt to the Kuiper belt \citep{2005Natur.435..462M}, and an explanation for the apparent late heavy bombardment (LHB) $3.8-4.1$ Gyr ago from lunar samples \citep{2005Natur.435..466G}. There are other models, and even extensions to the Nice model, but we focus on this one because of last point above. The fact that evidence exists of massive impacts of large asteroids $\sim$4 Gyr ago in the inner Solar System after the formation of the Earth--Moon system suggests another potential sourcing term for ETAs. \citet{2005Natur.435..466G} suggests impacts to the moon by asteroids as massive as $8\times10^{21}$ g. Assuming 2 g cm$^{-3}$, which is suggested in Figure 7 of \citet{2012P&SS...73...98C} for asteroids of this mass, this is an asteroid of $\sim200$ km in diameter. The escape velocity of the moon is $\sim2.4$ km s$^{-1}$, so it is feasible that LHB sourced inner Solar System Trojan asteroids including ETAs through impacts such as those suggested here, but also with other inner planets, moons, or minor planets. 

\indent We still stand by our assessment that if ETAs were sourced by LHB or earlier from the giant impact hypothesis or any other early Solar System mechanisms that were able to survive to the present arrangement of the Solar System, these types of ETAs would be considered ``primordial'' either way. Each of these occurred before the Solar System was 1 Gyr old leaving as much as 3.5 Gyr to survive to the present day. While previous work has shown ETA stability on the age of the Solar System is possible \citep{Cuk2012MNRAS.426.3051C, Zhuo2019A&A...622A..97Z}, our simulations point toward it being unlikely.

\indent Of the 11.2 million orbits initialized in the WHfast data set, 22,441 ETAs total survive a 1 Gyr integration. We showed in Figure \ref{fig:bootstrap} that increased integration extrapolated to more precise lifetimes shorter than the age of the Solar System (see Table \ref{table:fitparameters}). Extrapolation of the stretched exponential obtained from the full 1 Gyr of data predicts zero ETAs will remain in the simulation by 2.33 Gyr.

\indent To compare, there are roughly 20,000 near-Earth objects (NEOs) larger than 100m in the inner Solar System \citep{2011ApJ...743..156M,2014ApJ...792...30M}. This makes a good comparison for our surviving ETAs, since in \S\ref{subsec:yarkovsky} we argued that our simulation should be interpreted only to contain those larger than 100 meters because anything smaller would have been even more rapidly ejected due to the Yarkovsky effect. Primordial ETAs are unlikely to be more numerous than NEOs of the same size, which have a constant source in asteroids ejected from the main asteroid belt. Thus, we argue that zero ETAs in our simulation is indicative of zero primordial ETAs today\footnote{We will explore the potential of ground based surveys for both the surviving primordial ETA population as well as interloping ETAs like $2010 TK_7$ and $2020 XL_5$, for which we have collected a lot of information on in MEGASIM -- namely those that eject in time scales similar to the estimated lifetime of the known interloping ETAs.}.

\subsection{Extrapolating the stretched exponential fit}

\indent As we continue to run the simulations, we will be able to put the predicted zero time to the test by comparing the actual time when the simulations run out of asteroids to the predicted zero time. We note that should all of the simulations run out of asteroids, it does not directly indicate that the Solar System would run out of asteroids, but rather that with the current set of simulations the remaining asteroid fraction today would be less than one in 11.2 million of the initial population. This raises the question: is 11.2 million enough? Indeed, 11.2 million was picked for our own practical reasons more than anything. It is a large number, larger than the literature has seen by a wide margin, but also it fit on a manageable amount of nodes on our HPC to run them in entire batches to completion within our HPC wall-clock limit. In principle, one could simulation the settling of debris into the L4 or L5 points in a proto-planetary disk, or the settling of debris from later stage large impacts, but this is beyond the scope of our work, and we know of no such simulation in the literature.

\indent Finally, the zero time for the best fit stretched exponential (see Equation \ref{eq:stretchexp}) crosses zero because the constant term $C$ is negative. To obtain a fit at timescales of 10 kyr and 1 Gyr the constant factor is required. Without it equation \ref{eq:stretchexp} does not fit the simulation data well at both 10 kyr and 1 Gyr timescales, nor does it capture the steep downturn in remaining ETAs leading up to 1 Gyr. Without an additive constant an exponential has an asymptote on the x-axis never reaching zero. Which if taken at face value would imply an exponentially decreasing population would survive over arbitrarily long timescales. In practice a cut off should be made on the remaining ETA fraction, as below some fraction the likelihood of at least 1 remaining ETA becomes effectively zero. Where such a cut off should be made requires a constraint on the primordial mass that was initially bound to L4. As we mentioned in \S\ref{subsec:primordial} the primordial L4 mass is not well constrained by Solar System models.

\subsection{Conclusions}

\indent HPC at Lawrence Livermore National Laboratory has provided us the opportunity to build the largest ever simulation of ETAs initialized over the full Keplerian orbital parameter space with unprecedented fidelity. In this work we presented an analysis of the population of ETAs in regards to their lifetime as a population. Using the WHfast integrator the orbital data sets span an integration time of 1 Gyr for an initialized population of 11.2 million asteroids and all eight planets. We have continued integrating the simulation, but our results here stand for themselves. Further analysis is planned for our simulated data set. Coming papers will cover observability in current and future astronomical surveys such as the Zwicky Transient Facility and the Vera C. Rubin Observatory's Legacy Survey of Space and Time. We will also explore Earth impact likelihoods and the trajectories of quasi-stable interlopers like the two known ETAs. 

\indent Here we conclude our discussion with a brief list of the main findings of our analysis in the context of our literature review. 

\begin{itemize}
    \item We predict no ETAs can survive to the present day if those asteroids were initialized in the early Solar System.
    \item The fractional remainder of ETAs in our simulation over time follows a stretched exponential form. Extrapolating the remaining population of ETAs predicts zero by 2.33 Gyr.
    \item We performed an extensive review of stretch exponential functions in the literature, and we find similarities in those cases to ours here, and we related the physics of stretched exponentials to additive diffusion processes that are ultimately unable to relax in time before they are all ejected.
    \item Using a FFT analysis of the ejected ETAs and the planet's orbital dynamics, we showed that portions of the ejection pattern of ETAs aligns well with changes in Earth's orbital elements that cause the Milankovitch cycles that have been indicated in the climate record of Earth. The most important orbital variations seem to be in the semi-major axis, eccentricity and inclination of Earth.
    \item We showed that broad signals in the ejection pattern of the ETAs align with many resonances in all of the planet's orbital elements, which indicates the complexity of the physical process and also leads to the stretched exponential functional form and our interpretation. 
    \item We continue to integrate the simulations beyond 1 Gyr. When all asteroids have been ejected or if the functional form of the surviving ETAs changes our interpretation, we will follow up on our analysis. 
\end{itemize} 

\acknowledgments
\indent We thank Dr. Curt Struck for collaborative meetings discussing these results and his suggestion that the surviving ETA population may be best fit by a stretched exponential. We thank Drs. Michael Schneider and Caleb Miller for insight in theoretical underpinnings of our mathematical presentation here. We thank Dr. Eddie Schlafly for insightful discussions on studying the FFTs and presenting our results. We thank Dr. Will Dawson and Carmen Carrano for useful discussions and troubleshooting early on in the project. We also thank Noah Lifset and Ryan Dana for their earlier work and getting us started with \texttt{REBOUND}. Finally, we thank Dr. Renu Malhotra for sharing expertise and resources on ETAs that helped us focus our results. This work was performed under the auspices of the U.S. Department of Energy by Lawrence Livermore National Laboratory under Contract DE-AC52-07NA27344 and was supported by the LLNL-LDRD Program under Project 20-ERD-025.

\bibliography{main}

\begin{thebibliography}{}
\expandafter\ifx\csname natexlab\endcsname\relax\def\natexlab#1{#1}\fi
\providecommand{\url}[1]{\href{#1}{#1}}
\providecommand{\dodoi}[1]{doi:~\href{http://doi.org/#1}{\nolinkurl{#1}}}
\providecommand{\doeprint}[1]{\href{http://ascl.net/#1}{\nolinkurl{http://ascl.net/#1}}}
\providecommand{\doarXiv}[1]{\href{https://arxiv.org/abs/#1}{\nolinkurl{https://arxiv.org/abs/#1}}}

\bibitem[{{Almeida} {et~al.}(2009){Almeida}, {Peixinho}, \&
  {Correia}}]{2009A&A...508.1021A}
{Almeida}, A.~J.~C., {Peixinho}, N., \& {Correia}, A.~C.~M. 2009, \aap, 508,
  1021, \dodoi{10.1051/0004-6361/200911943}

\bibitem[{Alvarellos {et~al.}(2005)Alvarellos, Zahnle, Dobrovolskis, \&
  Hamill}]{alvarellos2005fates}
Alvarellos, J.~L., Zahnle, K.~J., Dobrovolskis, A.~R., \& Hamill, P. 2005,
  Icarus, 178, 104

\bibitem[{Alvarellos {et~al.}(2008)Alvarellos, Zahnle, Dobrovolskis, \&
  Hamill}]{alvarellos2008transfer}
---. 2008, Icarus, 194, 636

\bibitem[{Armstrong {et~al.}(2002)Armstrong, Wells, \&
  Gonzalez}]{armstrong2002rummaging}
Armstrong, J.~C., Wells, L.~E., \& Gonzalez, G. 2002, Icarus, 160, 183

\bibitem[{{Beekman}(2005)}]{2005JBAA..115..207B}
{Beekman}, G. 2005, Journal of the British Astronomical Association, 115, 207

\bibitem[{{Belbruno} \& {Gott}(2005)}]{2005AJ....129.1724B}
{Belbruno}, E., \& {Gott}, J.~Richard, I. 2005, \aj, 129, 1724,
  \dodoi{10.1086/427539}

\bibitem[{{Bershadskii}(2015)}]{2015arXiv151001909B}
{Bershadskii}, A. 2015, arXiv e-prints, arXiv:1510.01909.
\newblock \doarXiv{1510.01909}

\bibitem[{Bershadskii(2015)}]{bershadskii2015distributed}
Bershadskii, A. 2015, arXiv preprint arXiv:1512.08837

\bibitem[{Bershadskii(2016)}]{bershadskii2016buoyancy}
---. 2016, arXiv preprint arXiv:1608.00489

\bibitem[{{Boyanovsky} \& {Herring}(2019)}]{2019PhRvD.100b3531B}
{Boyanovsky}, D., \& {Herring}, N. 2019, \prd, 100, 023531,
  \dodoi{10.1103/PhysRevD.100.023531}

\bibitem[{Burns \& Gladman(1998)}]{burns1998dynamically}
Burns, J.~A., \& Gladman, B.~J. 1998, Planetary and space science, 46, 1401

\bibitem[{Bäckström {et~al.}(2014)Bäckström, Fischer, \& Boley}]{6951993}
Bäckström, T., Fischer, J., \& Boley, D. 2014, in 2014 22nd European Signal
  Processing Conference (EUSIPCO), 71--75

\bibitem[{{Cambioni} {et~al.}(2018){Cambioni}, {Malhotra}, {Hergenrother},
  {Rizk}, {Kidd}, {Drouet d'Aubigny}, {Chesley}, {Shelly}, {Christensen},
  {Farnocchia}, \& {Lauretta}}]{2018LPI....49.1149C}
{Cambioni}, S., {Malhotra}, R., {Hergenrother}, C.~W., {et~al.} 2018, in Lunar
  and Planetary Science Conference, Lunar and Planetary Science Conference,
  1149

\bibitem[{{Canup}(2012)}]{2012Sci...338.1052C}
{Canup}, R.~M. 2012, Science, 338, 1052, \dodoi{10.1126/science.1226073}

\bibitem[{{Carry}(2012)}]{2012P&SS...73...98C}
{Carry}, B. 2012, \planss, 73, 98, \dodoi{10.1016/j.pss.2012.03.009}

\bibitem[{Chambers {et~al.}(2002)Chambers, Quintana, Duncan, \&
  Lissauer}]{chambers2002symplectic}
Chambers, J.~E., Quintana, E.~V., Duncan, M.~J., \& Lissauer, J.~J. 2002, The
  Astronomical Journal, 123, 2884

\bibitem[{{Christou}(2019)}]{2019DDA....5010006C}
{Christou}, A. 2019, in AAS/Division of Dynamical Astronomy Meeting, Vol.~51,
  AAS/Division of Dynamical Astronomy Meeting, 100.06

\bibitem[{{Christou} \& {Georgakarakos}(2021)}]{2021MNRAS.507.1640C}
{Christou}, A.~A., \& {Georgakarakos}, N. 2021, \mnras, 507, 1640,
  \dodoi{10.1093/mnras/stab2223}

\bibitem[{{Collins} {et~al.}(2012){Collins}, {Kritsuk}, {Padoan}, {Li}, {Xu},
  {Ustyugov}, \& {Norman}}]{2012ApJ...750...13C}
{Collins}, D.~C., {Kritsuk}, A.~G., {Padoan}, P., {et~al.} 2012, \apj, 750, 13,
  \dodoi{10.1088/0004-637X/750/1/13}

\bibitem[{{Connors} {et~al.}(2000){Connors}, {Veillet}, {Wiegert}, {Innanen},
  \& {Mikkola}}]{2000DPS....32.1407C}
{Connors}, M., {Veillet}, C., {Wiegert}, P., {Innanen}, K., \& {Mikkola}, S.
  2000, in AAS/Division for Planetary Sciences Meeting Abstracts, Vol.~32,
  AAS/Division for Planetary Sciences Meeting Abstracts \#32, 14.07

\bibitem[{{Connors} {et~al.}(2011){Connors}, {Wiegert}, \&
  {Veillet}}]{2011Natur.475..481C}
{Connors}, M., {Wiegert}, P., \& {Veillet}, C. 2011, \nat, 475, 481,
  \dodoi{10.1038/nature10233}

\bibitem[{{{\'C}uk} {et~al.}(2012){{\'C}uk}, {Hamilton}, \&
  {Holman}}]{Cuk2012MNRAS.426.3051C}
{{\'C}uk}, M., {Hamilton}, D.~P., \& {Holman}, M.~J. 2012, \mnras, 426, 3051,
  \dodoi{10.1111/j.1365-2966.2012.21964.x}

\bibitem[{{de La Fuente Marcos} \& {de La Fuente
  Marcos}(2013)}]{2013MNRAS.432L..31D}
{de La Fuente Marcos}, C., \& {de La Fuente Marcos}, R. 2013, \mnras, 432, L31,
  \dodoi{10.1093/mnrasl/slt028}

\bibitem[{{de la Fuente Marcos} \& {de la Fuente
  Marcos}(2015)}]{2015MNRAS.453.1288D}
{de la Fuente Marcos}, C., \& {de la Fuente Marcos}, R. 2015, \mnras, 453,
  1288, \dodoi{10.1093/mnras/stv1725}

\bibitem[{{de la Fuente Marcos} \& {de la Fuente
  Marcos}(2017)}]{2017RNAAS...1....3D}
---. 2017, Research Notes of the American Astronomical Society, 1, 3,
  \dodoi{10.3847/2515-5172/aa95b5}

\bibitem[{{de la Fuente Marcos} \& {de la Fuente
  Marcos}(2021)}]{2021RNAAS...5...29D}
---. 2021, Research Notes of the American Astronomical Society, 5, 29,
  \dodoi{10.3847/2515-5172/abe6ad}

\bibitem[{Dobrovolskis {et~al.}(2000)Dobrovolskis, Alvarellos, \&
  Zahnle}]{dobrovolskis2000dynamical}
Dobrovolskis, A., Alvarellos, J., \& Zahnle, K. 2000, in AAS/Division for
  Planetary Sciences Meeting Abstracts\# 32, Vol.~32, 60--05

\bibitem[{Dobrovolskis {et~al.}(2007)Dobrovolskis, Alvarellos, \&
  Lissauer}]{DOBROVOLSKIS2007481}
Dobrovolskis, A.~R., Alvarellos, J.~L., \& Lissauer, J.~J. 2007, Icarus, 188,
  481, \dodoi{https://doi.org/10.1016/j.icarus.2006.11.024}

\bibitem[{Dobrovolskis \& Lissauer(2004)}]{dobrovolskis2004fate}
Dobrovolskis, A.~R., \& Lissauer, J.~J. 2004, Icarus, 169, 462

\bibitem[{Dones {et~al.}(1999)Dones, Gladman, Melosh, Tonks, Levison, \&
  Duncan}]{dones1999dynamical}
Dones, L., Gladman, B., Melosh, H., {et~al.} 1999, Icarus, 142, 509

\bibitem[{Dones {et~al.}(1996)Dones, Levison, \& Duncan}]{dones1996dynamical}
Dones, L., Levison, H., \& Duncan, M. 1996, in Completing the Inventory of the
  Solar System, Vol. 107, 233--244

\bibitem[{{Dvorak} {et~al.}(2012){Dvorak}, {Lhotka}, \&
  {Zhou}}]{2012A&A...541A.127D}
{Dvorak}, R., {Lhotka}, C., \& {Zhou}, L. 2012, \aap, 541, A127,
  \dodoi{10.1051/0004-6361/201118374}

\bibitem[{Eichelsbacher \& K{\"o}nig(2008)}]{10.1214/EJP.v13-539}
Eichelsbacher, P., \& K{\"o}nig, W. 2008, Electronic Journal of Probability,
  13, 1307 , \dodoi{10.1214/EJP.v13-539}

\bibitem[{Evans \& Tabachnik(1999)}]{evans1999possible}
Evans, N.~W., \& Tabachnik, S. 1999, Nature, 399, 41

\bibitem[{Farinella {et~al.}(1997)Farinella, Marzari, \&
  Matteoli}]{farinella1997disruption}
Farinella, P., Marzari, F., \& Matteoli, S. 1997, The Astronomical Journal,
  113, 2312

\bibitem[{{Gasperini} \& {Lolli}(2009)}]{2009PEPI..175..183G}
{Gasperini}, P., \& {Lolli}, B. 2009, Physics of the Earth and Planetary
  Interiors, 175, 183, \dodoi{10.1016/j.pepi.2009.03.011}

\bibitem[{{Gilmore} {et~al.}(2010){Gilmore}, {Kilmartin}, {Mauduit}, {Pforr},
  {Alvarez}, {Larsen}, {Mainzer}, {Wright}, {Bauer}, {Grav}, {Dailey},
  {Masiero}, {Cutri}, {McMillan}, {Walker}, {Miller}, {Miles}, {Roche}, \&
  {Foglia}}]{2010MPEC....T...45G}
{Gilmore}, A.~C., {Kilmartin}, P.~M., {Mauduit}, J., {et~al.} 2010, Minor
  Planet Electronic Circulars, 2010-T45

\bibitem[{{Giorgini} {et~al.}(1996){Giorgini}, {Yeomans}, {Chamberlin},
  {Chodas}, {Jacobson}, {Keesey}, {Lieske}, {Ostro}, {Standish}, \&
  {Wimberly}}]{1996DPS....28.2504G}
{Giorgini}, J.~D., {Yeomans}, D.~K., {Chamberlin}, A.~B., {et~al.} 1996, in
  AAS/Division for Planetary Sciences Meeting Abstracts, Vol.~28, AAS/Division
  for Planetary Sciences Meeting Abstracts \#28, 25.04

\bibitem[{Gladman(1997)}]{gladman1997destination}
Gladman, B. 1997, Icarus, 130, 228

\bibitem[{Gladman {et~al.}(2005)Gladman, Dones, Levison, \&
  Burns}]{gladman2005impact}
Gladman, B., Dones, L., Levison, H.~F., \& Burns, J.~A. 2005, Astrobiology, 5,
  483

\bibitem[{Gladman \& Duncan(1990)}]{gladman1990fates}
Gladman, B., \& Duncan, M. 1990, The Astronomical Journal, 100, 1680

\bibitem[{Gladman {et~al.}(2000)Gladman, Michel, \&
  Froeschl{\'e}}]{gladman2000near}
Gladman, B., Michel, P., \& Froeschl{\'e}, C. 2000, Icarus, 146, 176

\bibitem[{Gladman {et~al.}(1996)Gladman, Burns, Duncan, Lee, \&
  Levison}]{gladman1996exchange}
Gladman, B.~J., Burns, J.~A., Duncan, M., Lee, P., \& Levison, H.~F. 1996,
  Science, 271, 1387

\bibitem[{Gladman {et~al.}(1995)Gladman, Burns, Duncan, \&
  Levison}]{gladman1995dynamical}
Gladman, B.~J., Burns, J.~A., Duncan, M.~J., \& Levison, H.~F. 1995, Icarus,
  118, 302

\bibitem[{{Gomes} {et~al.}(2005){Gomes}, {Levison}, {Tsiganis}, \&
  {Morbidelli}}]{2005Natur.435..466G}
{Gomes}, R., {Levison}, H.~F., {Tsiganis}, K., \& {Morbidelli}, A. 2005, \nat,
  435, 466, \dodoi{10.1038/nature03676}

\bibitem[{Halsey \& Leibig(1991)}]{halsey1991stretched}
Halsey, T.~C., \& Leibig, M. 1991, Physical Review A, 43, 7087

\bibitem[{Hamilton \& Zeehandelaar(2006)}]{hamilton2006fate}
Hamilton, D.~P., \& Zeehandelaar, D. 2006, in AAS/Division of Dynamical
  Astronomy Meeting\# 37, Vol.~37, 13--02

\bibitem[{Hartmann {et~al.}(2000)Hartmann, Ryder, Dones, \&
  Grinspoon}]{hartmann2000time}
Hartmann, W., Ryder, G., Dones, L., \& Grinspoon, D. 2000, Origin of the Earth
  and Moon, 493

\bibitem[{{Hollis}(2022)}]{2022NatAs...6..178H}
{Hollis}, M. 2022, Nature Astronomy, 6, 178, \dodoi{10.1038/s41550-022-01615-0}

\bibitem[{Holman(1997)}]{holman1997possible}
Holman, M.~J. 1997, Nature, 387, 785

\bibitem[{Holman \& Wisdom(1993)}]{holman1993dynamical}
Holman, M.~J., \& Wisdom, J. 1993, The Astronomical Journal, 105, 1987

\bibitem[{{JeongAhn} \& {Malhotra}(2010)}]{2010DPS....42.1303J}
{JeongAhn}, Y., \& {Malhotra}, R. 2010, in AAS/Division for Planetary Sciences
  Meeting Abstracts, Vol.~42, AAS/Division for Planetary Sciences Meeting
  Abstracts \#42, 13.03

\bibitem[{Johnston(2006)}]{johnston2006stretched}
Johnston, D. 2006, Physical Review B, 74, 184430

\bibitem[{Johnston {et~al.}(2005)Johnston, Baek, Zong, Borsa, Schmalian, \&
  Kondo}]{johnston2005dynamics}
Johnston, D., Baek, S.-H., Zong, X., {et~al.} 2005, Physical review letters,
  95, 176408

\bibitem[{{Joyce} \& {Worrakitpoonpon}(2010)}]{2010JSMTE..10..012J}
{Joyce}, M., \& {Worrakitpoonpon}, T. 2010, Journal of Statistical Mechanics:
  Theory and Experiment, 2010, 10012, \dodoi{10.1088/1742-5468/2010/10/P10012}

\bibitem[{Kaps {et~al.}(2001)Kaps, Brando, Trinkl, B{\"u}ttgen, Loidl, Scheidt,
  Klemm, \& Horn}]{kaps2001heavy}
Kaps, H., Brando, M., Trinkl, W., {et~al.} 2001, Journal of Physics: Condensed
  Matter, 13, 8497

\bibitem[{{Kirkwood}(1891)}]{1891SidM...10..194K}
{Kirkwood}, D. 1891, Sidereal Messenger, 10, 194

\bibitem[{Laskar(1990)}]{laskar1990chaotic}
Laskar, J. 1990, Icarus, 88, 266

\bibitem[{{Laskar} {et~al.}(2011){Laskar}, {Fienga}, {Gastineau}, \&
  {Manche}}]{2011A&A...532A..89L}
{Laskar}, J., {Fienga}, A., {Gastineau}, M., \& {Manche}, H. 2011, \aap, 532,
  A89, \dodoi{10.1051/0004-6361/201116836}

\bibitem[{Levison \& Duncan(1994)}]{levison1994long}
Levison, H.~F., \& Duncan, M.~J. 1994, Icarus, 108, 18

\bibitem[{{Lifset} {et~al.}(2021){Lifset}, {Golovich}, {Green}, {Armstrong}, \&
  {Yeager}}]{2021AJ....161..282L}
{Lifset}, N., {Golovich}, N., {Green}, E., {Armstrong}, R., \& {Yeager}, T.
  2021, \aj, 161, 282, \dodoi{10.3847/1538-3881/abf7af}

\bibitem[{{Love}(2019)}]{2019PhDT.......109L}
{Love}, T. L.~M. 2019, PhD thesis, Technical University of Dresden, Germany

\bibitem[{{Mainzer} {et~al.}(2011){Mainzer}, {Grav}, {Bauer}, {Masiero},
  {McMillan}, {Cutri}, {Walker}, {Wright}, {Eisenhardt}, {Tholen}, {Spahr},
  {Jedicke}, {Denneau}, {DeBaun}, {Elsbury}, {Gautier}, {Gomillion}, {Hand},
  {Mo}, {Watkins}, {Wilkins}, {Bryngelson}, {Del Pino Molina}, {Desai},
  {G{\'o}mez Camus}, {Hidalgo}, {Konstantopoulos}, {Larsen}, {Maleszewski},
  {Malkan}, {Mauduit}, {Mullan}, {Olszewski}, {Pforr}, {Saro}, {Scotti}, \&
  {Wasserman}}]{2011ApJ...743..156M}
{Mainzer}, A., {Grav}, T., {Bauer}, J., {et~al.} 2011, \apj, 743, 156,
  \dodoi{10.1088/0004-637X/743/2/156}

\bibitem[{{Mainzer} {et~al.}(2014){Mainzer}, {Bauer}, {Cutri}, {Grav},
  {Masiero}, {Beck}, {Clarkson}, {Conrow}, {Dailey}, {Eisenhardt}, {Fabinsky},
  {Fajardo-Acosta}, {Fowler}, {Gelino}, {Grillmair}, {Heinrichsen}, {Kendall},
  {Kirkpatrick}, {Liu}, {Masci}, {McCallon}, {Nugent}, {Papin}, {Rice},
  {Royer}, {Ryan}, {Sevilla}, {Sonnett}, {Stevenson}, {Thompson}, {Wheelock},
  {Wiemer}, {Wittman}, {Wright}, \& {Yan}}]{2014ApJ...792...30M}
{Mainzer}, A., {Bauer}, J., {Cutri}, R.~M., {et~al.} 2014, \apj, 792, 30,
  \dodoi{10.1088/0004-637X/792/1/30}

\bibitem[{Malyshkin \& Tremaine(1999)}]{malyshkin1999keplerian}
Malyshkin, L., \& Tremaine, S. 1999, Icarus, 141, 341

\bibitem[{{Markwardt} {et~al.}(2020){Markwardt}, {Gerdes}, {Malhotra},
  {Becker}, {Hamilton}, \& {Adams}}]{2020MNRAS.492.6105M}
{Markwardt}, L., {Gerdes}, D.~W., {Malhotra}, R., {et~al.} 2020, \mnras, 492,
  6105, \dodoi{10.1093/mnras/staa232}

\bibitem[{{Marzari} \& {Scholl}(2013)}]{2013CeMDA.117...91M}
{Marzari}, F., \& {Scholl}, H. 2013, Celestial Mechanics and Dynamical
  Astronomy, 117, 91, \dodoi{10.1007/s10569-013-9478-7}

\bibitem[{Marzari {et~al.}(2002)Marzari, Tricarico, \&
  Scholl}]{marzari2002saturn}
Marzari, F., Tricarico, P., \& Scholl, H. 2002, The Astrophysical Journal, 579,
  905

\bibitem[{{Mignan}(2015)}]{2015GeoRL..42.9726M}
{Mignan}, A. 2015, \grl, 42, 9726, \dodoi{10.1002/2015GL066232}

\bibitem[{{Montesinos} {et~al.}(2020){Montesinos}, {Garrido-Deutelmoser},
  {Olofsson}, {Giuppone}, {Cuadra}, {Bayo}, {Sucerquia}, \&
  {Cuello}}]{2020A&A...642A.224M}
{Montesinos}, M., {Garrido-Deutelmoser}, J., {Olofsson}, J., {et~al.} 2020,
  \aap, 642, A224, \dodoi{10.1051/0004-6361/202038758}

\bibitem[{{Morbidelli} {et~al.}(2005){Morbidelli}, {Levison}, {Tsiganis}, \&
  {Gomes}}]{2005Natur.435..462M}
{Morbidelli}, A., {Levison}, H.~F., {Tsiganis}, K., \& {Gomes}, R. 2005, \nat,
  435, 462, \dodoi{10.1038/nature03540}

\bibitem[{Muller \& MacDonald(1997)}]{muller1997spectrum}
Muller, R.~A., \& MacDonald, G.~J. 1997, Proceedings of the national academy of
  sciences, 94, 8329

\bibitem[{{Murray} \& {Dermott}(1999)}]{1999ssd..book.....M}
{Murray}, C.~D., \& {Dermott}, S.~F. 1999, {Solar system dynamics} (Cambridge
  University Press)

\bibitem[{{Najafi} {et~al.}(2020){Najafi}, {Darooneh}, {Gheibi}, \&
  {Farhang}}]{2020ApJ...894...66N}
{Najafi}, A., {Darooneh}, A.~H., {Gheibi}, A., \& {Farhang}, N. 2020, \apj,
  894, 66, \dodoi{10.3847/1538-4357/ab8301}

\bibitem[{Nesvorn{\`y} {et~al.}(2003)Nesvorn{\`y}, Alvarellos, Dones, \&
  Levison}]{nesvorny2003orbital}
Nesvorn{\`y}, D., Alvarellos, J.~L., Dones, L., \& Levison, H.~F. 2003, The
  Astronomical Journal, 126, 398

\bibitem[{Nesvorn{\`y} \& Dones(2002)}]{nesvorny2002long}
Nesvorn{\`y}, D., \& Dones, L. 2002, Icarus, 160, 271

\bibitem[{{Nobili} {et~al.}(1989){Nobili}, {Milani}, \&
  {Carpino}}]{1989A&A...210..313N}
{Nobili}, A.~M., {Milani}, A., \& {Carpino}, M. 1989, \aap, 210, 313

\bibitem[{Palmer {et~al.}(1984)Palmer, Stein, Abrahams, \&
  Anderson}]{palmer1984models}
Palmer, R.~G., Stein, D.~L., Abrahams, E., \& Anderson, P.~W. 1984, Physical
  Review Letters, 53, 958

\bibitem[{Peterson(1989)}]{peterson1989time}
Peterson, I. 1989, Science News, 135, 157

\bibitem[{Phillips(1996)}]{phillips1996stretched}
Phillips, J. 1996, Reports on Progress in Physics, 59, 1133

\bibitem[{Quintana {et~al.}(2002)Quintana, Lissauer, Chambers, \&
  Duncan}]{quintana2002terrestrial}
Quintana, E.~V., Lissauer, J.~J., Chambers, J.~E., \& Duncan, M.~J. 2002, The
  Astrophysical Journal, 576, 982

\bibitem[{{Rein} \& {Liu}(2012)}]{rebound}
{Rein}, H., \& {Liu}, S.~F. 2012, \aap, 537, A128,
  \dodoi{10.1051/0004-6361/201118085}

\bibitem[{{Rein} \& {Spiegel}(2015)}]{reboundias15}
{Rein}, H., \& {Spiegel}, D.~S. 2015, \mnras, 446, 1424,
  \dodoi{10.1093/mnras/stu2164}

\bibitem[{{Rein} \& {Tamayo}(2015)}]{reboundwhfast}
{Rein}, H., \& {Tamayo}, D. 2015, \mnras, 452, 376,
  \dodoi{10.1093/mnras/stv1257}

\bibitem[{Robutel \& Laskar(2001)}]{robutel2001frequency}
Robutel, P., \& Laskar, J. 2001, Icarus, 152, 4

\bibitem[{{Santana-Ros} {et~al.}(2022){Santana-Ros}, {Micheli}, {Faggioli},
  {Cennamo}, {Devog{\`e}le}, {Alvarez-Candal}, {Oszkiewicz}, {Ram{\'\i}rez},
  {Liu}, {Benavidez}, {Campo Bagatin}, {Christensen}, {Wainscoat}, {Weryk},
  {Fraga}, {Brice{\~n}o}, \& {Conversi}}]{2022NatCo..13..447S}
{Santana-Ros}, T., {Micheli}, M., {Faggioli}, L., {et~al.} 2022, Nature
  Communications, 13, 447, \dodoi{10.1038/s41467-022-27988-4}

\bibitem[{{Sato} {et~al.}(2019){Sato}, {Nakamura}, {Goto}, {Kumagai},
  {Nagahama}, {Minoura}, {Zhao}, {Heslop}, \& {Roberts}}]{2019E&PSL.520...94S}
{Sato}, T., {Nakamura}, N., {Goto}, K., {et~al.} 2019, Earth and Planetary
  Science Letters, 520, 94, \dodoi{10.1016/j.epsl.2019.05.028}

\bibitem[{Scher(1991)}]{scher1991time}
Scher, H. 1991, Physics Today, 26

\bibitem[{{Szebehely}(1967)}]{1967AJ.....72....7S}
{Szebehely}, V. 1967, \aj, 72, 7, \dodoi{10.1086/110195}

\bibitem[{Tabachnik \& Evans(2000)}]{tabachnik2000asteroids}
Tabachnik, S., \& Evans, N. 2000, Monthly Notices of the Royal Astronomical
  Society, 319, 63

\bibitem[{{Todd} {et~al.}(2012){Todd}, {Tanga}, {Coward}, \&
  {Zadnik}}]{2012MNRAS.420L..28T}
{Todd}, M., {Tanga}, P., {Coward}, D.~M., \& {Zadnik}, M.~G. 2012, \mnras, 420,
  L28, \dodoi{10.1111/j.1745-3933.2011.01186.x}

\bibitem[{Trinkl {et~al.}(2000)Trinkl, B{\"u}ttgen, Kaps, Loidl, Klemm, \&
  Horn}]{trinkl20007}
Trinkl, W., B{\"u}ttgen, N., Kaps, H., {et~al.} 2000, Physical Review B, 62,
  1793

\bibitem[{{Tsiganis} {et~al.}(2005){Tsiganis}, {Gomes}, {Morbidelli}, \&
  {Levison}}]{2005Natur.435..459T}
{Tsiganis}, K., {Gomes}, R., {Morbidelli}, A., \& {Levison}, H.~F. 2005, \nat,
  435, 459, \dodoi{10.1038/nature03539}

\bibitem[{{Whiteley} \& {Tholen}(1998)}]{1998Icar..136..154W}
{Whiteley}, R.~J., \& {Tholen}, D.~J. 1998, \icarus, 136, 154,
  \dodoi{10.1006/icar.1998.5995}

\bibitem[{{Wiegert} {et~al.}(2000){Wiegert}, {Innanen}, \&
  {Mikkola}}]{2000Icar..145...33W}
{Wiegert}, P., {Innanen}, K., \& {Mikkola}, S. 2000, \icarus, 145, 33,
  \dodoi{10.1006/icar.2000.6339}

\bibitem[{Wilson \& Toumi(2005)}]{wilson2005fundamental}
Wilson, P., \& Toumi, R. 2005, Geophysical Research Letters, 32

\bibitem[{{Wisdom} \& {Holman}(1991)}]{wh}
{Wisdom}, J., \& {Holman}, M. 1991, The Astronomical Journal, 102, 1528,
  \dodoi{10.1086/115978}

\bibitem[{Yeager \& Golovich(2022)}]{Yeager_2022}
Yeager, T., \& Golovich, N. 2022, Research Notes of the {AAS}, 6, 68,
  \dodoi{10.3847/2515-5172/ac62da}

\bibitem[{{Yizhaq} {et~al.}(2020){Yizhaq}, {Xu}, \&
  {Ashkenazy}}]{2020E&PSL.54416373Y}
{Yizhaq}, H., {Xu}, Z., \& {Ashkenazy}, Y. 2020, Earth and Planetary Science
  Letters, 544, 116373, \dodoi{10.1016/j.epsl.2020.116373}

\bibitem[{{Yoshida} \& {Nakamura}(2005)}]{2005AJ....130.2900Y}
{Yoshida}, F., \& {Nakamura}, T. 2005, \aj, 130, 2900, \dodoi{10.1086/497571}

\bibitem[{{Yoshikawa} {et~al.}(2018){Yoshikawa}, {Tsuda}, {Watanabe}, {Tanaka},
  {Nakazawa}, {Terui}, {Saiki}, \& {Hayabusa2 Project
  Team}}]{2018LPI....49.1771Y}
{Yoshikawa}, M., {Tsuda}, Y., {Watanabe}, S., {et~al.} 2018, in Lunar and
  Planetary Science Conference, Lunar and Planetary Science Conference, 1771

\bibitem[{{Zahnle} {et~al.}(2007){Zahnle}, {Alvarellos}, {Dobrovolskis}, \&
  {Hamill}}]{zahnle2007transfer}
{Zahnle}, K., {Alvarellos}, J., {Dobrovolskis}, A., \& {Hamill}, P. 2007, in
  38th Annual Lunar and Planetary Science Conference, Lunar and Planetary
  Science Conference, 2001

\bibitem[{Zeehandelaar \& Hamilton(2005)}]{zeehandelaar2005material}
Zeehandelaar, D., \& Hamilton, D. 2005, in AAS/Division for Planetary Sciences
  Meeting Abstracts\# 37, Vol.~37, 61--18

\bibitem[{{Zhou} {et~al.}(2019){Zhou}, {Xu}, {Zhou}, {Dvorak}, \&
  {Li}}]{Zhuo2019A&A...622A..97Z}
{Zhou}, L., {Xu}, Y.-B., {Zhou}, L.-Y., {Dvorak}, R., \& {Li}, J. 2019, \aap,
  622, A97, \dodoi{10.1051/0004-6361/201834026}

\end{thebibliography}
\bibliographystyle{aasjournal}



\label{lastpage}
\end{document}